\documentclass{elsart} 

\usepackage{graphicx}
\usepackage{amssymb}
\usepackage{xr} 

\begin{document}
\date{accepted in Physica D}


\begin{frontmatter}

\title{Nonlinear dynamics of waves and modulated waves in 1D thermocapillary flows.\\
I: General presentation and periodic solutions}

\author{Nicolas Garnier\thanksref{atlanta}},
\author{Arnaud Chiffaudel\corauthref{ac}},
\ead{arnaud.chiffaudel@cea.fr}
\author{Fran\c{c}ois Daviaud}\\
and
\author{Arnaud Prigent}

\address{Groupe Instabilit\'es et Turbulence,
	Service de Physique de l'Etat Condens\'e, \\
        Direction des Sciences de la Mati\`ere, CEA Saclay, CNRS URA 2464, \\
        B\^at. 772, Orme des Merisiers, 91191 Gif-sur-Yvette, France}

\thanks[atlanta]{Present address : Center for Nonlinear Science, 
		Georgia Institute of Technology GA 30332-0430.
		{\tt garnier@cns.physics.gatech.edu} }
\corauth[ac]{Corresponding author} 

\begin{abstract}

We present experimental results on hydrothermal traveling-waves dynamics
in long and narrow 1D channels. The onset of primary traveling-wave
patterns is briefly presented for different fluid heights and for
annular or bounded channels, {\em i.e.}, within periodic or non-periodic boundary
conditions. For periodic boundary conditions, by increasing the control
parameter or changing the discrete mean-wavenumber of the waves, we
produce modulated waves patterns. These patterns range from stable
periodic phase-solutions, due to supercritical Eckhaus instability, to
spatio-temporal defect-chaos involving traveling holes and/or
counter-propagating-waves competition, {\em i.e.}, traveling sources and
sinks. The transition from non-linearly saturated Eckhaus modulations to
transient pattern-breaks by traveling holes and spatio-temporal defects
is documented. Our observations are presented in the framework of
coupled complex Ginzburg-Landau equations with additional fourth and
fifth order terms which account for the reflection  
symmetry breaking at high wave-amplitude far from onset. The second part of this
paper~\cite{part2} extends this study to spatially non-periodic patterns
observed in both annular and bounded channel.

\end{abstract}


\end{frontmatter}

\tableofcontents
\newpage


\label{sec:intro}
\section*{Introduction}

The transition to spatio-temporal chaos in extended non-linear systems
remains still today an active field of research. Many studies have been
devoted to this problem in the case of stationary spatial patterns and
more recently in the case of oscillatory instabilities and non-linear
traveling-waves patterns. For example, Rayleigh-B\'enard
convection~\cite{davbon90} and directional viscous
fingering~\cite{rabmic90} in quasi one-dimensional experiments have
revealed a transition to spatio-temporal chaos via spatio-temporal
intermittency. On the other hand, non-linear traveling-waves have
exhibited a fascinating variety of behaviors and patterns. Wave systems
have been studied in binary-fluid convection (subcritical traveling
waves bifurcation)~\cite{kol92,kol92b,baxeat92}, oscillatory instability
in low Prandtl number convection~\cite{janpum92,legjan92}, oscillatory
rotating convection~\cite{liueck99}, cylinder wake~\cite{lewpro95},
Taylor dean vortices. \cite{botcad98,botmut00}

Among these different waves systems, thermocapillary flows and in
particular hydrothermal waves
\cite{davvin93,mukchi98,garchi01,garchi02,garchi01c,burmuk01} or hot
wire waves \cite{vindub92,vindub97,alvhec97,paswes01} appear as a very
simple tool, owing to their supercritical bifurcation. They correspond
to the first instability of a thin liquid layer with a free surface
subjected to a horizontal temperature gradient~\cite{smidav83}.
Different experimental configurations have been used to study those
traveling waves: rectangular cells with different aspect
ratios~\cite{davvin93,rilnei98,pelbur99,burmuk01,garchi01}, annular
cells~\cite{schmol92,mukchi98}, cylindrical
cells~\cite{garnier00,garchi01c} and linear hot wire under the surface
of a liquid~\cite{vindub92,vindub97,alvhec97,paswes01}. 

Hydrothermal waves provide very interesting systems of traveling waves
which can be modeled by envelope equations such as the complex
Ginzburg-Landau equation (CGL)~\cite{crohoh93,arakra01}. The CGL
equation, which describes the large-scale modulations of the bifurcated
solutions near oscillatory instabilities, has been extensively studied
as shown in a recent review~\cite{arakra01}. This situation is due both
to its relevance to many experimental systems, even far from threshold,
and to the variety of spatio-temporal chaos regimes is exhibits. Theoretical
studies and numerical simulations have indeed revealed regimes of phase
turbulence~\cite{shrpum92,torfra97}, amplitude turbulence
\cite{shrpum92} related to localized defects solutions such as the
Bekki-Nozaki or homoclonic holes~\cite{nozbek85,hecke98,burcha99} and,
more recently, modulated amplitude
waves~\cite{bruzim00,brutor01a,brutor01b}. 

\subsection*{Geometries and symetries}

We study two different one-dimensional traveling-wave systems: one in an
annular geometry and one in a rectangular geometry. The annular setup
has periodic boundary conditions and the rectangular setup is finite,
bounded with poorly reflecting boundaries. Both are spatially extended.
Our results are presented in two companion articles. The present paper
begins with a general introduction to the dynamics of hydrothermal waves
and its modeling. Then it presents and discuss experimental data in the
form of uniform or modulated wave-patterns obtained in the annular cell
and corresponding to periodic solutions of the problem. A second paper
\cite{part2} is devoted to non-periodic and non-uniform patterns, {\em
i.e.}, to patterns either observed in the rectangular bounded channel or
observed in the annular channel in cases where the galilean invariance
is broken by the presence of fixed defects. The distinction we make
between those two classes of experiments is motivated by important
differences between the observed wave-patterns. Whereas the present
paper I presents results that can be directly connected with numerics or
analytics in periodic or infinite systems, this will not the case in
paper II, where we will emphasize the convective and absolute regimes
and transitions for instabilities. In the following, we will refer to
the two papers as I and II and will sometime refer to paper II to
provide the reader a broader view on wave-dynamics in actual
experiments, where the galilean invariance usually doesn't hold,
which leads to qualitative and quantitative differences from the work
presented in the present paper.

The pattern in annular geometry undergoes a supercritical Eckhaus
instability in the form of stable traveling modulations~\cite{mukchi98}.
We detail this in the present paper, together with other modulated
patterns. We also present realizations of spatio-temporal chaos in
our periodic wave-system. We mention similar patterns obtained
in rectangular geometry, though their accurate description is given in II.

\subsection*{Outline of the article}

This article is segmented as follows: section~\ref{sec:hydro} presents
the experimental setups, the main characteristics of hydrothermal waves
and CGL modeling. This section is of general interest for the readers of
both papers I and II. Section~\ref{sec:annulus} is then devoted to the
description of modulated traveling waves in the annulus for a medium
fluid height where extensive quantitative measurements have been
realized. Section~\ref{sec:annulus_bis} presents additional data about
modulated wave patterns observed for smaller fluid heights. Finally, in
section~\ref{sec:annulus_disc}, we discuss the modulated-wave regimes and
develop a comparison with theoretical and numerical solutions.

\section{Hydrothermal waves}
\label{sec:hydro}

\subsection{Experimental setups}
\label{sec:setup}

Hydrothermal waves have been studied in two different 1D geometries:
annular corresponding to periodic boundary conditions \cite{mukchi98}
and rectangular corresponding to finite bounded ones
\cite{garchi01,garchi02}. Both geometries consist of a channel with a
glass bottom and vertical copper walls filled with a thin layer of
silicon oil of viscosity $\nu = 0.65$cSt and Prandtl number $P = 10$
(see~\cite{burmuk01} for characteristics of the fluid). The fluid
surface is free and a Plexiglas plate is inserted a few millimeters
above the surface of the fluid to reduce evaporation.

\begin{figure}
\begin{center}
\includegraphics[height=7.5cm]{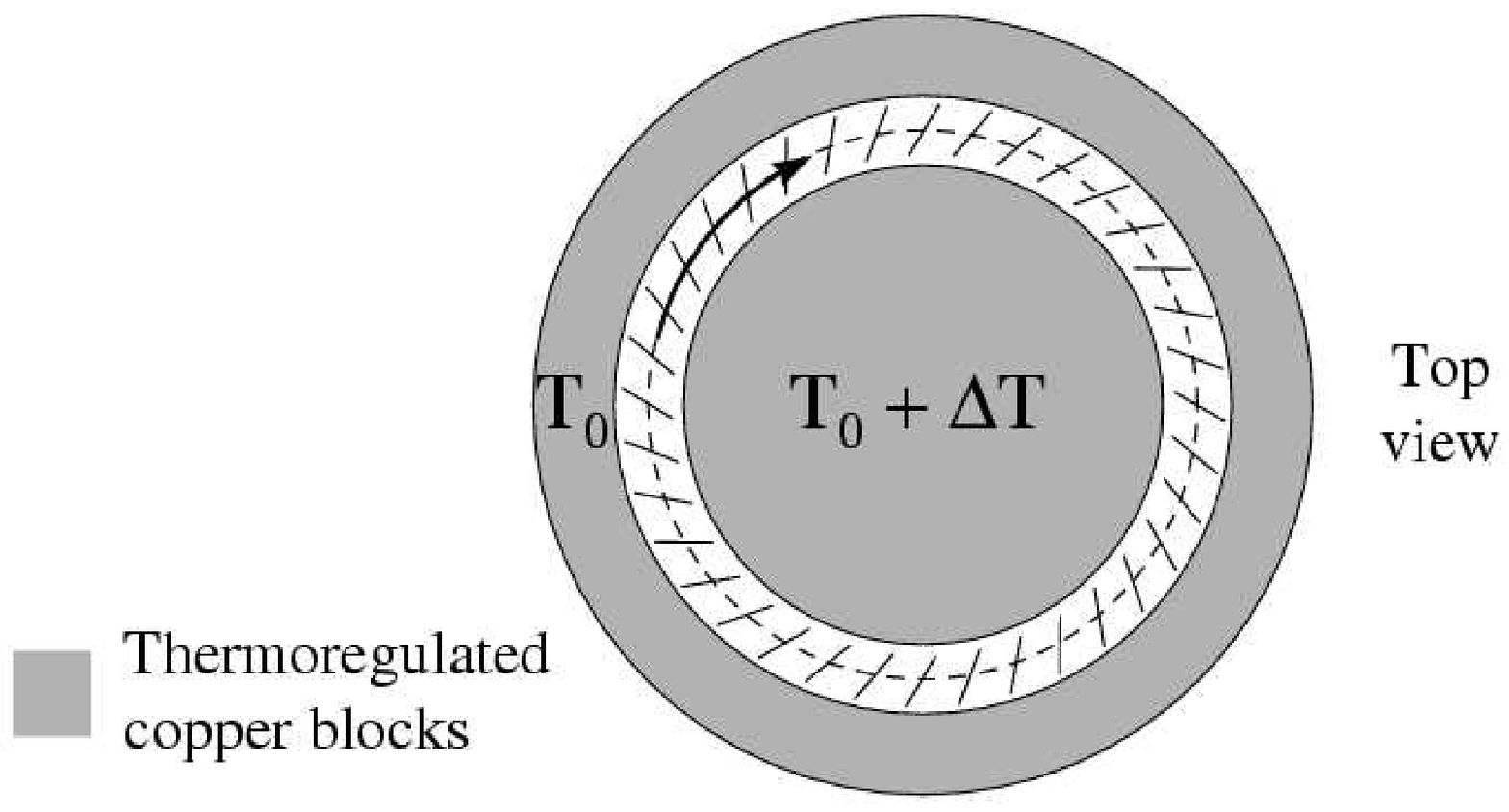}
\includegraphics[width=10cm]{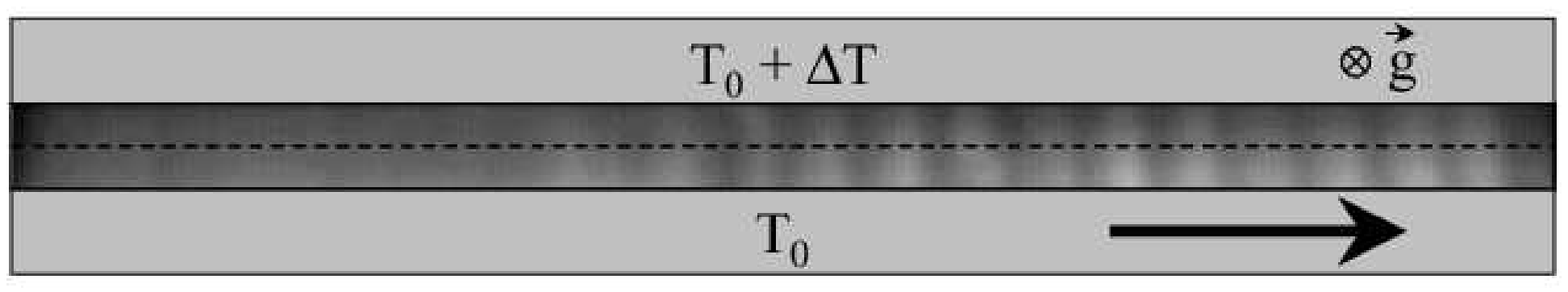}
\end{center}

\caption{Schematics of the annular periodic channel (top) and
rectangular bounded channel (bottom). Traveling hydrothermal waves are
represented by a sketch in the annulus and by a photograph in the
rectangle. Dashed lines shows the acquisition circle and line
respectively.}

\label{fig:schema_ann}
\label{fig:schema_rect}
\end{figure}

The annular channel (Fig.~\ref{fig:schema_ann}) is $10$mm wide and its
mean radius is $R = 80$~mm~\cite{mukchi98}. The fluid height $h$ is varied
between $0.6$ and $3$~mm. The perimeter is $L_p=503$~mm which
corresponds to an aspect ratio $\Gamma = 2\pi R/h = 300$ for $h =
1.7$~mm. The outer copper wall is cooled by a thermo-regulated fluid
circulation at $293$~K while the inner copper block is heated
electrically.

The rectangular channel (Fig.~\ref{fig:schema_rect}) is $10$\,mm wide
and $250$\,mm long~\cite{davvin93,burmuk01}. Plexiglas blocks are
inserted in the channel to reduce the length to $L_b = 180$~mm, {\em
i.e.}, aspect ratio $\Gamma = L_b/h \simeq 100$; this allows us to view
the whole cell through a $200$~mm diameter lens. The copper walls are
thermo-regulated by fluid circulations.

Let's define our notation for the channel length: the channel length
will be noted $L_{\rm p}$ for the periodic (annular) channel and $L_{\rm
b}$ for the bounded (rectangular) channel; without subscript, $L$ will
concern the current channel, and $L^*$ the non-dimensional length
$L/\xi_0$, where $\xi_0$ stands for the correlation length of the
system. $\xi_0$ depends of $h$ but not of the channel boundaries (see
section~\ref{sec:model}).

Most experiments reported in this paper are performed around $h=1.7$~mm.
Otherwise, the height will be noted in text and figure captions.
Maintaining the height as constant as possible is a challenge due to
evaporation and thermocapillary flow of the oil on the vertical sides of
the channels. Height decrease rate is typically $0.1$~mm/h in the
rectangular channel and $0.01$~mm/h in the annular channel where more
attention has been paid to reduce thermocapillary side-flowing. In some
case, the decrease rate has been as low as $0.01$~mm/day. Quantitative
data presented below will concern measurements in the range $h=(1.70 \pm
0.05)$~mm for the annulus and $h=(1.7 \pm 0.1)$~mm for the rectangle.

In both cases thermocouples allow accurate measurements of the
temperature difference $\Delta T$ across the channel, typically
established with a $\pm 15$mK stability. 

\begin{figure}
\begin{center}
\includegraphics[height=4cm]{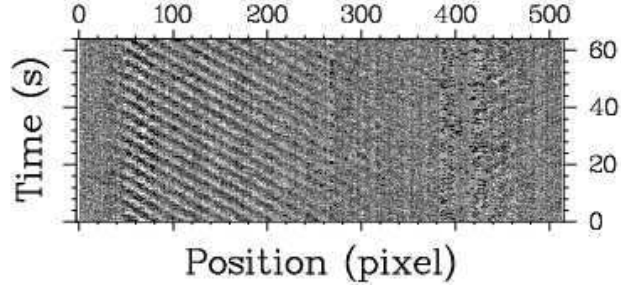}
\end{center}

\caption{Exemple of unprocessed spatio-temporal diagram of a left
traveling wave in the rectangular channel close to onset ($\Delta
T=3.8$~K). Only a few periods are presented while usual acquisitions last
several hours. The cell is shorter than the acquisition line.}

\label{fig:dst_rect}
\end{figure}

Convective patterns are observed through the glass bottom by
shadowgraphy. A parallel vertical white light beam crosses the container
from top to bottom and forms a horizontal picture on a screen, mainly
due to temperature gradients in the
fluid~\cite{garnier00,Settles:01,MayFel:01}. This configuration insures
a low contrast and thus a linear response to the waves.
Fig.~\ref{fig:schema_rect} shows a typical photograph of the
wave-pattern in the rectangular channel. Images are digitized with a CCD
camera over $512 \times 512$ or $768 \times 512$ pixels. Spatio-temporal
diagrams of 512 data points are then extracted from a line (rectangular
geometry) or a circle (annular geometry) and plotted along time
(Fig.~\ref{fig:dst_rect}). To extract the pattern behavior from the
spatio-temporal diagrams, space and time Fourier transforms and complex
demodulation techniques are used. They allow a determination of the
local amplitude of the waves, their wavenumber and frequency as
detailled in section~\ref{sec:demod}.

\subsection{hydrothermal waves}
\label{sec:hydrothermal}

The existence of hydrothermal waves in a fluid layer subjected to a
horizontal temperature gradient has been predicted on the basis of a
linear stability analysis by Smith and Davis \cite{smidav83}, and
studied numerically in conditions close to our experimental situation by
Mercier and Normand \cite{mernor96}. Hydrothermal waves have been
detected and characterized in several
experiments~\cite{davvin93,burmuk01,rilnei98,pelbur99,sch92}. In the
following, we recall some of their main features obtained in
quasi-one-dimensional geometries. More information can be found in
Ref.~\cite{burmuk01}.

Given all the physical properties of the fluid, the two parameters which
control our experimental system are $h$, the height of fluid in the
cell, and $\Delta T$, the horizontal temperature difference between the
two walls. The associated dimensionless parameters are:

\begin{itemize}
\item the Marangoni number: $Ma=\gamma \Delta T h / \rho \kappa \nu$
\item the dynamic Bond number $Bo= \rho g \alpha h^2/\gamma $  
\item the capillary number $Ca =\gamma \Delta T/\sigma$
\item the static Bond number $Bd= \rho g h^2 / \sigma$
\end{itemize}

\noindent where $g$ is the gravitational acceleration, $\alpha$ the
thermal expansion coefficient, $\kappa$ the thermal diffusivity, $\rho$
the density of the fluid, $\nu$ the kinematic viscosity, $\sigma$ the
surface tension and $\gamma =-(\partial\sigma/\partial T)$. The exact
characteristics of the waves depend strongly on these parameters and on
the aspect ratios of the containers~\cite{burmuk01,rilnei98,pelgar00}.
For the hydrothermal waves instability considered here, only $Ma$ and
$Bo$ are relevant. Moreover, in practice, the experiments reported in
this paper are performed with a constant width (10 mm) and for some
different fluid depths. Most of the quantitative results concerns
$h=1.7$~m. In the following, we will use $\Delta T$ as our control
parameter for simplicity. 
 
\begin{figure}
\begin{center} \includegraphics[width=10cm]{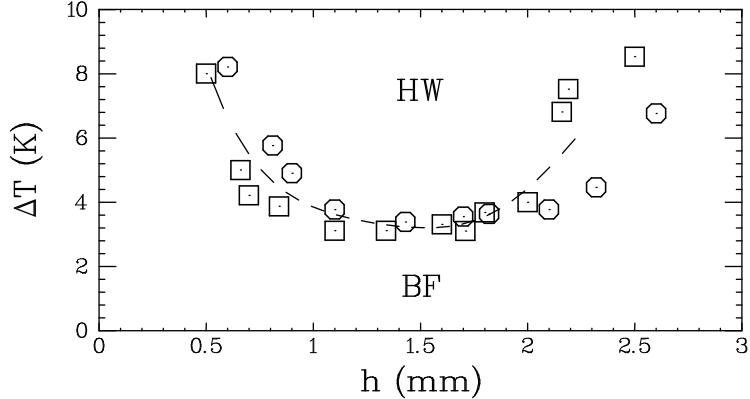} 
\end{center}

\caption{Stability diagram: critical temperature difference $\Delta T_c$
vs. height of liquid $h$ for a $10$~mm width channel. This graph
combines data obtained in rectangular ($\circ$) and annular ($\Box$)
geometry. BF and HW respectively refer to Basic Flow and Hydrothermal
Waves. These data ---except for $h=1.7$~mm, see
Fig.~\ref{fig:A_vs_DT}--- correspond to the lowest $\Delta T$ where
waves are visually seen to occupy the whole cell. These data roughly
give the same tendency, schematized by the dashed line. }


\label{StabDiag}
\end{figure}

As soon as $\Delta T \ne 0$, a convective flow is created. The thermal
gradient across the cell induces a surface tension gradient on the free
surface of the fluid. Due to Marangoni effect ($\gamma > 0$), this
gradient generates a surface flow towards the cold side, with a bottom
recirculation: the basic flow (BF) is a long annular (resp.
longitudinal) roll in the annular (resp. rectangular) case. Please note
that this basic flow is not due to an instability. However, above a
given threshold $\Delta T_c$, and for small depth layers $h < h_r$
($h_r=2.7$~mm for a $10$~mm width channel), this flow becomes unstable
with respect to oblique traveling waves (TW) which propagate along the
channel (Fig.~\ref{fig:schema_rect}), {\em i.e.}, the roll axis
\cite{burmuk01}. Above $h_r$, stationary rolls parallel to the gradient
are observed. $\Delta T_c$ depends on $h$: when increasing $h$, $\Delta
T_c$ decreases from small $h$ towards a minimum and then increases up to
$h_r$ (Fig.~\ref{StabDiag}). As soon as the threshold is crossed, two
waves which propagate in opposite direction along the roll axis, and
towards the hot wall appear in the container, separated by a source. The
source shape as well as the characteristics of the waves and their
evolution with time and $\Delta T$ above the threshold depend of $h$. 
 
\begin{figure}
\begin{center} \includegraphics[width=10cm]{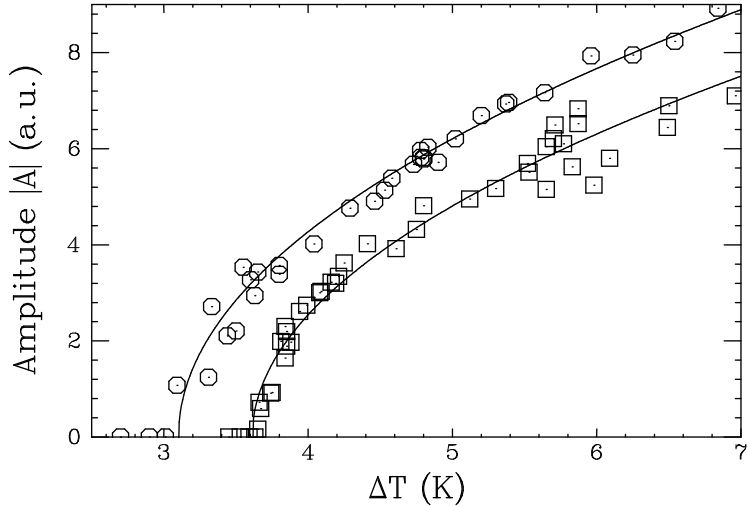}
\end{center}

\caption{Amplitude of the hydrothermal waves vs. $\Delta T$ for $h=1.7$
mm and $10$~mm width: experimental data in the annulus with $L_{\rm
p}=503$ mm ($\circ$) and in the rectangle with $L_{\rm b}=180$ mm
($\Box$). The annulus data are for patterns with mean wavenumber around
the critical wavenumber $k_{\rm c}$, {\em i.e.}, for $k=54(2\pi/L_{\rm p})$
and $k=55(2\pi/L_{\rm p})$. Solid lines represent fits $A \propto
\left(\Delta T-\Delta T_{\rm c}\right)^{1/2}$, $\Delta T_{\rm c}=3.1$K
for the annulus (convective threshold) and $A \propto \left(\Delta
T-\Delta T_{\rm a}\right)^{1/2}$, $\Delta T_{\rm a}=3.65$K for the
rectangle (absolute threshold \cite{garchi01}).}

\label{fig:A_vs_DT}
\end{figure}

The waves appear via a $I_{\rm o}$ bifurcation \cite{crohoh93}: as for a
supercritical Hopf bifurcation ---but for a spatially extended system---
the frequency is finite at threshold and the amplitude of the waves
behaves as $(\Delta T - \Delta T_c)^{1/2}$ \cite{garchi01,burmuk01}. In
this paper, we present for the first time the critical amplitude
behaviors of supercritical traveling waves in both type of cells:
bounded or periodic (Fig.~\ref{fig:A_vs_DT}). These results are in
agreement with the linearity of the shadowgraphic response of our
optical system. Please note the difference between the two thresholds
presented in Fig.~\ref{fig:A_vs_DT}: it is due to the difference between
the convective and the absolute threshold. It is carefully discussed in
the companion paper II. Please note also that finite size effects of
order $(2\pi/L)^2$ can be neglected in both channel because of large
spatial extensions.~\cite{garchi01,pelgar00}

In the stability domain of the waves, two types of sources have been
evidenced in the rectangular channel~\cite{burmuk01}. For larger
heights, the source is a line and generally evolves towards one end of
the container leaving a single wave whereas for smaller heights, the
source looks like a point and emits a circular wave which becomes almost
planar far from the source in both directions~\cite{burmuk01,garchi01c}.
In the periodic annular channel the two types of sources are also
observed, but only during transients; sources and sinks are both
unstable near onset always leaving a single wave; this is detailed in
section~\ref{sec:sources} of II. In the following, we report behaviors
obtained for small heights, but the width of the two geometries is small
(10 mm) and the waves behavior can still be considered as one-dimensional.

\subsection{Complex Ginzburg-Landau envelope equation modeling}
\label{sec:model}

The experimental shadowgraphic signal $\theta_r(x,t)$ is basically a
space- and time-periodic field (Fig.~\ref{fig:dst_rect}). Let's describe
high frequencies with Fourier modes and low-frequency dynamics
by two slowly varying amplitudes A and B:

\begin{eqnarray}
\theta_r(x,t) = A(X,T)\ &&\exp i(\omega_{\rm c} t-k_{\rm c} x) \nonumber \\
              \mbox{} + &&B(X,T)\ \exp i(\omega_{\rm c} t+k_{\rm c} x) + c.c. + ...
\end{eqnarray}

where $\omega_{\rm c}$ is the critical frequency, $k_{\rm c}$ is the
critical wavenumber, c.c. stands for complex conjugate and the dots for
harmonics. Near threshold, these slowly varying amplitudes may be
described by a system of two complex Ginzburg-Landau (CGL) equations:

\begin{eqnarray}
\tau_0 (A_T+s A_X) = \epsilon(1+ic_0) A &+& \xi_0^2 (1+ic_1)A_{XX} \nonumber \\
          \mbox{} &-& g(1+ic_2) |A|^2A - g(\lambda+i\mu)|B|^2A     \nonumber \\
\tau_0 (B_T-s B_X) = \epsilon(1+ic_0) B &+& \xi_0^2 (1+ic_1)B_{XX} \nonumber \\
          \mbox{} &-& g(1+ic_2) |B|^2B - g(\lambda+i\mu)|A|^2B
\label{eq:cgl}
\end{eqnarray}

where $\epsilon = \Delta T/\Delta T_{\rm c}-1$ is the non-dimensional
control parameter, $\tau_0$ is the characteristic time scale, $\xi_0$ is
the correlation length, $s$ is the group velocity, $g$ is a real
amplitude scaling factor ($g>0$ in our supercritical system), $c_0$
stands for a first order dependence of the wave-frequency with
$\epsilon$ and $c_1$, $c_2$, $\lambda$ and $\mu$ are the real
non-dimensional CGL coefficients. Since traveling waves are always
selected against standing waves, we know that $\lambda$ is bigger than
unity. Whereas dimensioning is necessary for experimental data modeling,
we will often use the non-dimensional form to discuss physical
properties in a simplified framework: $c_0$ is then set to $0$;
$\tau_0$, $\xi_0$, $g$, and even $\epsilon$ are set to $1$.

We wish to emphasize that the above definition of $\epsilon$ involves
$\Delta T_{\rm c}$, {\em i.e.}, the critical temperature difference.
This value corresponds to the onset of linear instability, {\em i.e.}, the
onset for {\em convectively unstable} waves.

CGL model equation describe correctly most non-linear wave systems such
as low-Prandtl-number oscillatory instability
\cite{crowil89,janpum92,legjan92}, binary fluid convection
\cite{kol92,kol92b}, convection with rotation \cite{liueck99}, cylinder
wake \cite{lewpro95} and so on. A single equation is enough for systems
with broken $x \mapsto -x$ symmetry, {\em i.e.}, when a single wave is
present. In our hydrothermal wave experiment, both equations for $A$ and
$B$ are required, except for some simple patterns high above onset (see
section~\ref{sec:UHWCGL}).

\subsection{Experimental demodulation technique and CGL modeling}
\label{sec:demod}

Following the usual framework of nonlinear patterns, we try to
extract from the spatio-temporal images quantities which could be
directly written in a nonlinear model equation such as CGL. For that
purpose, the real shadowgraphic data $\theta_r(x,t)$, related to the
thermal field, can be written: 

\begin{equation} \theta_r(x,t) = \theta_c(x,t) + c.c. + ... \end{equation} 

where $\theta_c$ is called the complexified signal, c.c. stands for the
complex conjugate and the dots for harmonics. Variables $x$ and $t$ are
the laboratory space and time. In order to demodulate the signal, we
apply Hilbert transform. Real data are complexified by the three
following operations \cite{crowil89,mukchi98}: a Fourier transform of
$\theta_r$ in $x$ or $t$, a wide band-pass filtering around the positive
fundamental frequency in Fourier space and then the inverse Fourier
transform; this creates the complex signal $\theta_c$. In this complex
signal, the right- and left-propagating waves data are mixed:

\begin{eqnarray} 
&\theta_c(x,t) &= r(x,t) + l(x,t)
\end{eqnarray} 

In order to separate $r(x,t)$ and $l(x,t)$, spatial filtering that
select positive and negative wavenumbers are performed. We get:

\begin{eqnarray} 
&r(x,t) &= A(X,T)\ \exp i(\omega_{\rm c} t - k_{\rm c} x) \nonumber \\ 
&l(x,t) &= B(X,T)\ \exp i(\omega_{\rm c} t + k_{\rm c} x) 
\label{eq:wavedef}
\end{eqnarray} 

where $A(X,T)$ and $B(X,T)$ are complex amplitudes depending of the slow
variables $X$ and $T$, which correspond exactly to the slowly varying
envelopes described by amplitude equations such as coupled CGL models.
In an experimental situation, the critical frequency $\omega_{\rm c}$
and wavenumber $k_{\rm c}$ need first to be measured. So our
demodulation technique decomposes the signals as:

\begin{eqnarray}
r(x,t) &=& |A(X,T)|\ \exp i \varphi_{\rm r}(x,t) \nonumber \\
l(x,t) &=& |B(X,T)|\ \exp i \varphi_{\rm l}(x,t) 
\end{eqnarray} 

where $\varphi_{\rm r,l}$ are two fast-varying phases, rotating at the
experimental signal frequency, and also containing slow-varying
modulations, {\em i.e.}, the phases $\Phi_{\rm r}(X,T)$ and $\Phi_{\rm
l}(X,T)$ of $A(X,T)$ and $B(X,T)$. In practice, the complex demodulation
of each shadowgraphic data set results in six spatio-temporal images:

\begin{itemize}

\item two moduli: $|A(X,T)|$ and $|B(X,T)|$, {\em i.e.}, the local and
instantaneous amplitudes of the waves.

\item two gradients of each right- and left-propagating phases
$\varphi_{\rm r}$ and $\varphi_{\rm l}$:
\begin{eqnarray} 
\omega &=& \partial_t \varphi_{\rm r,l} = \omega_{\rm c} + \partial_T \Phi_{\rm r,l}(X,T) \nonumber \\
     k &=& \partial_x \varphi_{\rm r,l} =      k_{\rm c} + \partial_X \Phi_{\rm r,l}(X,T)
\end{eqnarray}
{\em i.e.}, the local and instantaneous frequencies $\omega$ and
wavenumbers $k$ of the right and left wave-patterns.

\end{itemize}

Finally, let's remember that Fourier transform is bijective when it is
applied to infinite or periodical signals only. So, we choose to process
space-periodic spatio-temporal images by first filtering them in space,
and then in time. On the contrary, images from the rectangular bounded
box experiment are first filtered in time, in order to benefit from the
sharpness of time spectra obtained after long-time data-acquisition, and
then in space to separate right and left waves. Both techniques will
result in different signs to $\omega_{\rm c}$ and $k_{\rm c}$ which need
to be adapted to the chosen wave description (Eq.~\ref{eq:wavedef}).

In the following, $\omega$ and $k$ will refer to the local
phase-gradients. Their mean values will generally be given to
characterize the experimental patterns.

\subsection{Uniform hydrothermal waves (UHW) in annular geometry
and higher order CGL (HOCGL) modeling}
\label{sec:UHWCGL}
\label{sec:HOCGL}

\begin{figure}
\begin{center}
\includegraphics[height=13cm]{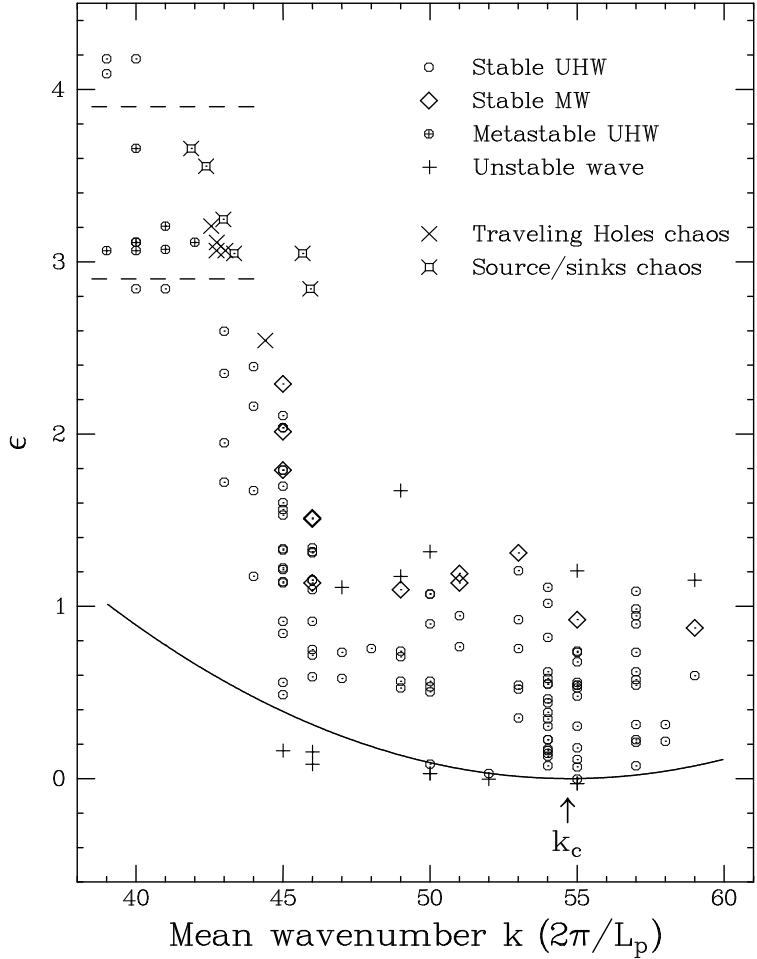}
\end{center}

\caption{Stability diagram in the $(\epsilon,k)$ space for the annular
experiment at fluid height $h=1.7$~mm. This diagram presents stable
uniform hydrothermal waves (UHW) observed along experimental runs
($\circ$). These UHW are characterized by their spatially-uniform and
temporally-constant wave amplitude, frequency and wavenumber $k$. The
plain line is the marginal stability curve. For high $\epsilon$ and $k$
closed to $k_c$, stable modulated waves (MW) are observed in a narrow
region ($\diamond$), which correspond to non-linearly saturated Eckhaus
instability patterns. Unstable states above MW as well as below the
marginal stability curve are shown by plusses ($+$): those states are
unstable, and relax either to steady flow ($\epsilon<0$) or to UHW with
different wavenumber $k$ ($\epsilon>0$). The $2.9 \lesssim \epsilon
\lesssim 3.9$ domain, delimited by dashed lines, is characterized by
spatio-temporal chaos and temporal intermittency: metastable low-$k$-UHW
($\oplus$) and disordered states with traveling holes ($\times$) and/or
traveling source/sink ($\times$ with small square) take place
alternatively over very slow time-scales (hours). These dynamical
regimes are presented in sections \ref{sec:annulus} and
\ref{sec:annulus_bis}. }

\label{fig:diagstab1}
\end{figure}

Performing the experiment in an annular channel is the easiest way to
approach ideal theoretical solutions. In this case simple solutions
---with $|A|$, $|B|$, $\omega$ and $k$ uniform in space and time--- are
observed. In bounded rectangular channels, end conditions imposes
spatial variations to at least the wave amplitude. This case will be
exposed in II. In the annulus, because of periodicity of the boundary
condition, a restriction occur: the mean wavenumber has to be an integer
in units of $2\pi/L_{\rm p}$. The reduced wavenumber or mean phase
gradient

\begin{equation}
q=\xi_0 \, (k-k_{\rm c})
\end{equation}

is thus a discrete quantity. For a given value of the control parameter,
different discrete wavenumbers may be stable during different
experimental runs. In fact the large aspect ratio leads to typically 50
wavelengths for $h=1.7$~mm. Driving the experiment back and forth from
onset to disorder at high $\Delta T$ allows to successively explore
different values of the wavenumber from $k=39$ to $60(2\pi/L_{\rm p})$.
One should note that these states are {\em stricto sensu} metastable
states since several integer $k$ are observed for a given $\epsilon$: we
will label them Uniform Hydrothermal Waves (UHW). The resulting
stability diagram is presented in Fig.~\ref{fig:diagstab1}. In the
central region (closed circles), we report uniform hydrothermal waves
(UHW). This region appears to be limited by the Eckhaus \cite{Eck:65}
secondary instability not only, as usual, for lowering $\epsilon$ but
also for high increasing $\epsilon$ \cite{mukchi98} in the central
wavenumber region for $k \gtrsim 45(2\pi/L_{\rm p})$. The dynamics of
the secondary instability will be presented in section~\ref{sec:annulus}
and discussed in section~\ref{sec:annulus_disc}. From the UHW
observation we get the wave amplitude $A$ and frequency $\omega=2\pi f$
as functions of $\epsilon$ and $k$. The coupling between right $A$ and
left $B$ waves results in a single wave pattern and we will consider
only a single CGL equation for $A$, setting $B=0$.

\begin{figure}
\begin{center}
\includegraphics[width=10cm]{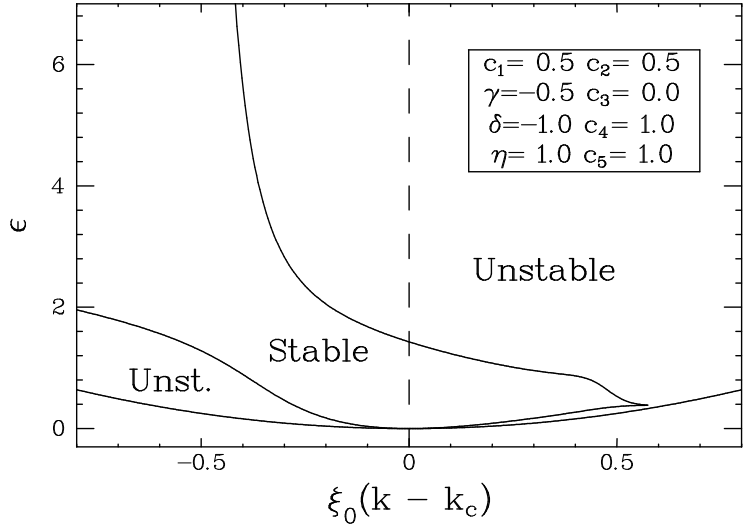}
\end{center}

\caption{Analytical computation of the Eckhaus instability limit in the
$(\epsilon,q)$ plane for HOCGL model equation~(\ref{eq:hocgl}) for
$c_1=0.5$, $c_2=0.5$, $\gamma=-0.5$, $c_3=0$, $\delta=-1$, $c_4=1$,
$\eta=1$, $c_5=1$. The set of coefficients has been chosen to
qualitatively follow the shape of the experimental diagram: please note
how the classical parabola is distorted and how the region of uniform
hydrothermal waves (UHW) is limited for high $\epsilon$ in the central
wavenumber band around $k_{\rm c}$. Due to the higher order terms (HOT),
the diagram looses the $q \mapsto -q$ symmetry for non-zero $\epsilon$.
The curves drawn here represents the Eckhaus small-wavenumber
instability limit, but we also verified that the system is stable
against modulations at any finite wavenumber~\cite{matvol93}. }

\label{fig:diagHOCGL}
\end{figure}

From the experimental data $A=A(\epsilon,k)$ and
$\omega=\omega(\epsilon,k)$ we tried to fit most coefficients of CGL
Eq.~(\ref{eq:cgl}). The results, which deserve careful quantitative
presentation and will be published in details elsewhere, are
qualitatively summarized below. For example, the marginal stability
curve is obtained by extrapolating the $|A(\epsilon)|^2$ curve to
$\epsilon=0$ for each fixed integer value of $k$. We show, from UHW data
only, that the CGL model is valid only in the close vicinity of the wave
onset. But the full stability region, up to $\epsilon \lesssim 2.5$, can
be described by introducing higher order terms (HOT)~\cite{legmen99},
scaling as $\epsilon^{4/2}$ and $\epsilon^{5/2}$ instead of the regular
$\epsilon^{3/2}$, and including the amplitude gradient $\partial_X A$
\cite{eckioo89}. The CGL is then modified to an higher order equation
(HOCGL):

\begin{eqnarray}
A_T+s A_X = \epsilon A + (1+ic_1)A_{XX} &-& (1+ic_2) |A|^2A \nonumber \\
&+& (\gamma +ic_3) \partial_X(|A|^2)A \nonumber \\
&+& (\delta +ic_4) |A|^2 A_X \nonumber \\
&+& (\eta   +ic_5) |A_X|^2 A, 
\label{eq:hocgl}
\end{eqnarray}

written here in its non-dimensional form. The first two additional terms
contains all the possible fourth order dependence in $\epsilon^{1/2}$.
The last one has been chosen among fifth order terms for its relevance
to describe experimental results. Several other fifth order terms may of
course be written. Fourth order terms break the $x \mapsto -x$ symmetry
of the problem. This seems, at first sight, to disagree with the basic
symmetry hypothesis which leads to the derivation of CGL envelope
equation. In fact, these terms play a role only for higher values
$\epsilon$ ---higher values of $A$--- when the propagation direction
itself is already responsible for the symmetry breaking. Close to onset,
these terms are negligible because of their higher order: HOCGL is then
equivalent to CGL. So, even far above onset, HOCGL equation appears to
be the right equation to model the dynamics and stability of UHW
solutions. As a consequence, the Eckhaus modulational instability
appears also for high $\epsilon$, and the border of the Eckhaus stable
domain is not symmetrical with respect to $(q \mapsto -q)$ in the
wavenumber space. Such asymmetrical stability domain shape is
encountered in other experiments, e.g., rotating disk
flow~\cite{schleg98}. The measurement of the coefficients of
Eq.~\ref{eq:hocgl} is still under progress. It is supported mainly by
UHW data $A(\epsilon,k)$ and $\omega(\epsilon,k)$ and basic properties
of Eckhaus unstable modulated waves (section~\ref{sec:annulus}). An
example of stability diagram for HOCGL is shown in
Fig.~\ref{fig:diagHOCGL}. This example has no uniform solutions above a
finite $\epsilon$ for $k$ close to $k_{\rm c}$; this leads to a
wavenumber selection process for increasing $\epsilon$~\cite{legmen99}.
For other coefficients sets, the upper curve may limit the whole
wavenumber band so the system behaves as being Benjamin-Feir unstable
\cite{benfei67} above finite $\epsilon$.

Section~\ref{sec:annulus} and \ref{sec:annulus_bis} are devoted to the
presentation of various experimental examples of modulated
wave-patterns, {\em i.e.}, patterns involving the Eckhaus instability
mechanism or other secondary instability mechanisms. Powerful
theoretical and numerical analysis of such patterns have been reported
recently, introducing the concept of modulated amplitude waves (MAWs)
\cite{bruzim00,brutor01a,brutor01b}, that will be presented in
section~\ref{sec:annulus_disc}. The authors use a single CGL equation
which has only three degrees of freedom, {\em i.e.}, the real
coefficients $c_1$ and $c_2$ and the mean wavenumber $q$ \cite{qnu}.
This is much less than our experimental system which uses eight
coefficients. Is a quantitative comparison between the experimental and
numerical \cite{bruzim00,brutor01a,brutor01b} solution feasible? A naive
conjecture would be that, for a given $\epsilon$, the system is locally
equivalent to a CGL model with appropriate $c_1(\epsilon)$ and
$c_2(\epsilon)$ (differing from the HOCGL constant $c_1$ and $c_2$). We
believe this conjecture is wrong, because even for constant given
$\epsilon$, the wavenumber dependence of the UHW amplitude
$A(\epsilon,q)$ differs from the well known
$A(\epsilon,q)=\sqrt{\epsilon-q^2}$: for high $\epsilon \gtrsim
1$, the amplitude is minimum in the center of the stable
wavenumber-band, and maximum at its boundaries! Quantitative comparisons
will thus require a minimal care when modeling the UHW properties with
non-symmetrical HOT in the model equation. However, despite this
complexity to quantitatively describe UHW solutions with a suitable
non-linear model, we will show how the experimental modulated waves and
the numerically known coherent structures called MAWs resemble on their shape,
profiles and dynamics (section~\ref{sec:annulus_disc}). 

Finally, let's note that UHW do not exist in the bounded channel (II)
because the amplitude should vanish at the boundaries, but quasi-UHW are
observed far above onset and below the secondary instability onset.

\subsection{Experimentally known CGL coefficients}
\label{sec:coeffs}

While theoretical or numerical research on CGL depends only on the value
of $c_1$ and $c_2$ (and $s$ in bounded domains), experimental work
starts with the necessity to reduce dimensional data, {\em i.e.}, to measure
$\omega_{\rm c}$, $k_{\rm c}$, $\Delta T_c$ (or $\epsilon$), $\tau_0$,
$\xi_0$, $s$ and $g$. For $h=1.7$~mm, we obtain~\cite{garchi01}:

\begin{eqnarray}
&&\omega_{\rm c} = 0.237             \, {\rm Hz}         \nonumber \\
&&k_{\rm c}      = (0.684 \pm 0.003) \, {\rm mm}^{-1}
                 = 54.75             \, (2\pi/L_{\rm p})
                 = 19.6              \, (2\pi/L_{\rm b}) \nonumber \\
&&\Delta T_c     = (3.1   \pm 0.1)   \, {\rm K}          \nonumber \\
&&\tau_0         = (5     \pm 1)     \, {\rm s}          \nonumber \\
&&\xi_0          = (5.1   \pm 0.3)   \, {\rm mm}         \nonumber \\
&&s              = (0.895 \pm 0.01)  \, {\rm mm.s}^{-1} 
\label{eq:coeff-s}
\end{eqnarray}

The status of $g$ is particular: it is just an amplitude unit,
converting arbitrary gray levels in non-dimensional units and depends on
the settings of the optical shadowgraphic device. In the annular device,
it is measured with a $10^{-2}$ relative accuracy.

Then $c_0$, $c_1$, $c_2$,..., $\lambda$ and $\mu$ may be measured mostly
by comparison with CGL solutions dynamics. In fact, the need to extend
CGL by HOT has increased the complexity of the process. While some
coefficients are easy to estimate, the most wanted $c_1$ and $c_2$ come
at the end of the process with large error bars. Quantitative
information may be released only in part: most coefficients are of order
of $1$ in absolute value, $c_1$ is small ($|c_1| \lesssim 0.5$)
\cite{garchi01} and the real coupling coefficient, measured in
transients with a new method \cite{caschi01}, is 

\begin{equation}
\lambda = 1.36 \pm 0.2
\label{lambda} 
\end{equation}

just bigger than unity. We have yet no way to estimate $\mu$. The
small value of $\lambda$ denotes a moderate destructive interaction
between right- and left-propagating waves compared, for example, to
oscillatory instability in Argon where $\lambda=2$ \cite{crowil89}.
Although standing hydrothermal waves have never been observed in
one-dimensional systems, complex states involving both right and left
waves are common in the rectangular cell, close to onset, and in the
annular cell, for chaotic regimes at large $\epsilon$: once the
amplitude of the major wave is attenuated ---and whatever the cause---
the opposite wave is much less damped and starts growing.

\section{Modulated traveling waves: 
                          general presentation at moderate fluid height}
\label{sec:annulus}

Simple wave patterns corresponding to basic Stokes plane-waves solutions
of CGL models in periodic boundary conditions have been briefly
presented in section~\ref{sec:UHWCGL}. The present section describes
modulated traveling waves observed in our annular channel. These
solutions result from secondary instabilities. As far as the data
reported in this paper are produced in narrow channels and are thus
one-dimensional, secondary instabilities may only be modulational, {\em
i.e.}, developing along the propagation axis. This includes Eckhaus
instability \cite{Eck:65}, Benjamin-Feir instability \cite{benfei67},
and excludes, e.g., zigzag instability. For steady patterns as
Rayleigh-B\'enard convection, modulational instabilities are subcritical
and not saturated by non-linear terms: once the pattern is unstable, a
modulation appears which leads to the apparition or to the annihilation
of a wavelength ---e.g., a roll pair \cite{lowgol85,krazim85}. However,
for traveling wave patterns, it is known that the development of the
modulation may be supercritical
\cite{fauve87,janpum92,mukchi98,brutor01b}, {\em i.e}, saturated by the
non-linearities, thus leading to stable modulated waves. It may also
be subcritical \cite{kol92,baxeat92,liueck99} just as in the steady
case. Modulated waves are characterized by oscillations of their
wavenumber, frequency and amplitude that travels in space and time at
the hydrothermal wave group velocity. Using radio transmission language,
a modulated wave may be described as a low frequency and long wavelength
signal ---the slowly varying complex amplitude $A(X,T)$--- combined with
a carrier wave ---the fast varying $\exp i(\omega_{\rm c}t - k_{\rm
c}x)$ (section~\ref{sec:model}).

\subsection{Supercritical Eckhaus instability generates modulated waves 
            near $k=k_{\rm c}$}

This section is devoted to the study of the stability of traveling waves
as a function of $\epsilon$ for $k$ close to $k_{\rm c}$, {\em i.e.}, in
the central band of the stability diagram (Fig.~\ref{fig:diagstab1}).
The complete study has been performed for $h=1.7$~mm.

\subsubsection{Damped modulated waves below Eckhaus onset}
\label{sec:mwdamped}

\begin{figure}
\begin{center}
\includegraphics[width=7cm]{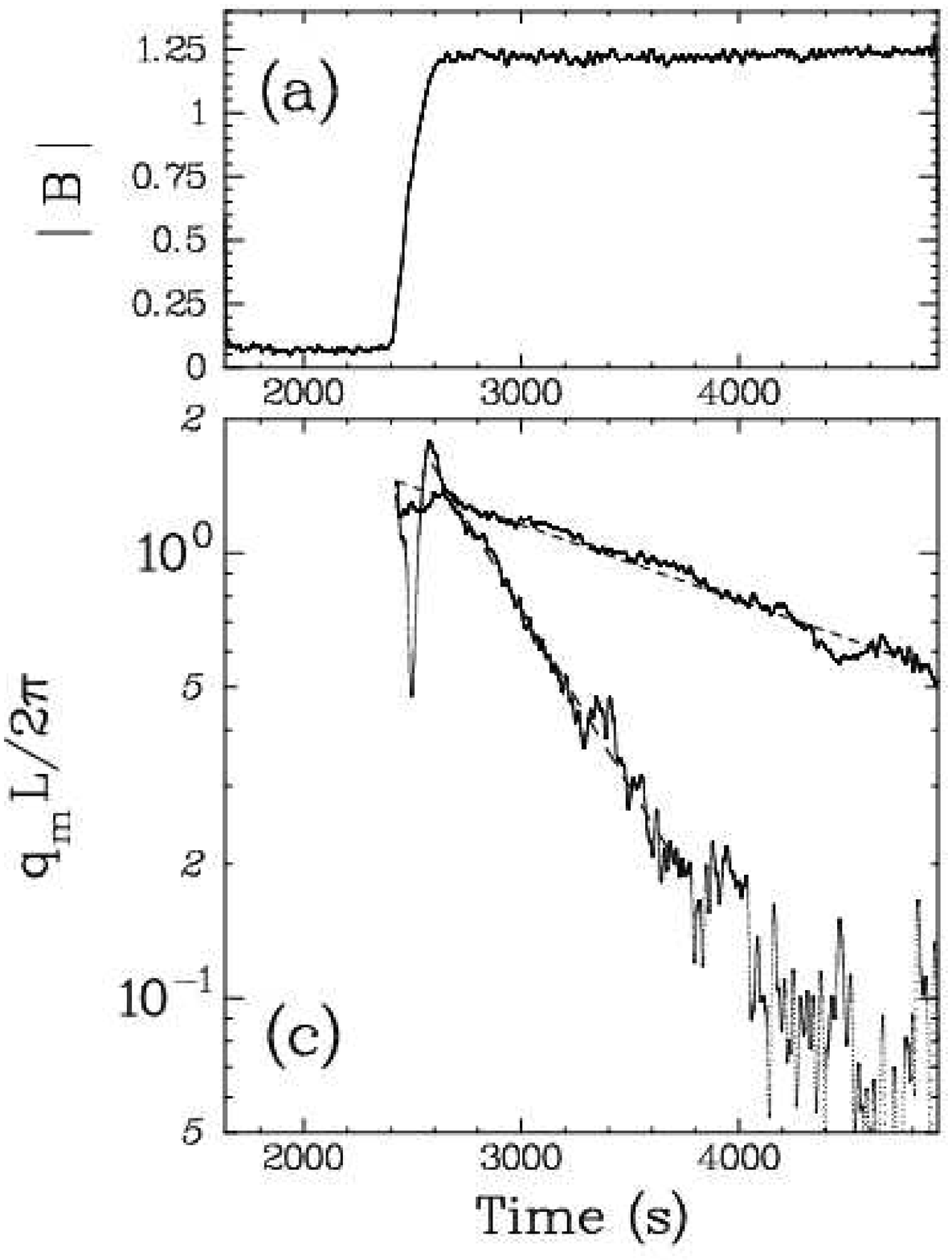}%
\includegraphics[width=7cm]{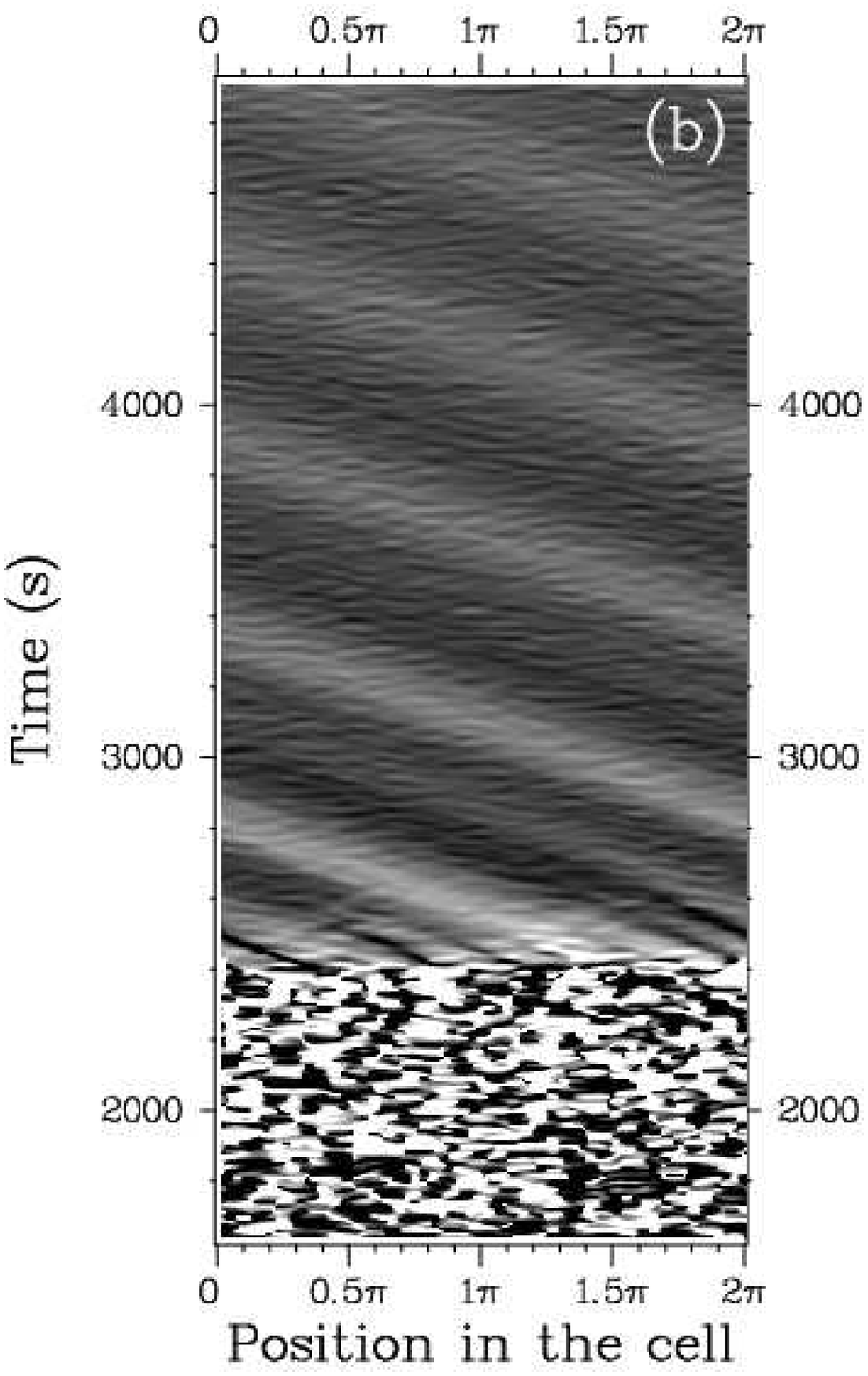} 
\end{center}

\caption{Apparition of modulated waves at hydrothermal wave onset
produced by a fast increase of $\epsilon$ from a small negative value
below onset to $\epsilon=0.16$ ($\Delta T= 3.60$~K) near $t=2250$~s.
(a):~Mean amplitude $|B|(T)|$ of the carrier left-traveling wave showing
the control parameter change.
(b):~Spatio-temporal diagram of the phase gradient $\partial_X \Phi$ of
the left-traveling wave. 
(c):~Temporal evolution of the wavenumber-modulation amplitude $q_{\rm
m}(T)$ extracted from phase-gradient data (b) by Hilbert transform. The
upper curve show the $K_{\rm M}=2\pi/L_{\rm p}$ fundamental mode of the
modulation and the lower curve corresponds to the first spatial harmonic
at $K_{\rm M}=2 \cdot (2\pi/L_{\rm p})$. The dashed lines are fits of
the exponential decays. }

\label{fig:mwdamped}
\end{figure}

Let start with a description of an experiment from linearly stable flow
at negative $\epsilon$. When crossing threshold, a wave-pattern appears.
A transient regime with competing counter-propagating wave is generally
observed. This transient leads to a single right- or left-traveling wave
as $\epsilon$ is increased. Once the single wave orbits along the
channel, we notice that it is always a modulated wave. 

An example is presented in Fig.~\ref{fig:mwdamped}. The modulated
hydrothermal wave appears near $t=2250$~s after $\epsilon$ has been
rapidly increased from a small negative value to $\epsilon=0.16$
($\Delta T= 3.60$~K). 

Fig.~\ref{fig:mwdamped}a presents the mean amplitude $|B(T)|$ of the
left-traveling wave, {\em i.e.}, the relevant order parameter for
this transition. The right-traveling wave $A$ exists only during a few
hundred seconds at the begining of the transient. The transient itself
is presented in details in paper II (section~\ref{sec:sources}): the
amplitudes for both waves and the detail of the competition are
illustrated on a spatio-temporal diagram
(Fig.~\ref{fig:st_competition} of II).

Fig.~\ref{fig:mwdamped}b presents the local wavenumber $\partial_X \Phi$
of the left-traveling dominant wave. The phase is not defined until the
amplitude of the wave becomes finite. Then, the wavenumber sets up
around the mean value $k=53(2\pi/L_{\rm p})$, {\em i.e.}, close to
$k_c$. This picture shows the local variations of the wavenumber, which
propagate at the group velocity and decay with time. This characterizes
a modulated wave (MW) decaying toward a uniform hydrothermal wave (UHW).
The modulation wavenumber is $K_{\rm M}=2\pi/L_{\rm p}$, {\em i.e.}, the
smallest possible wavenumber in the cell. The modulation appears to be
damped: using a second Hilbert transform applied on $\partial_X \Phi$
(Fig.~\ref{fig:mwdamped}b), one can extract the amplitude $q_{\rm m}(T)$
of the local wavenumber modulation:

\begin{equation}
\partial_X \Phi(X,T) = q_{\rm m}(T) \ \exp i(\Omega_{\rm M} T - K_{\rm M} X) + ...
\label{eq:defq}
\end{equation}

This modulation amplitude is presented in Fig.~\ref{fig:mwdamped}c. The
$K_{\rm M}=2\pi/L_{\rm p}$ fundamental mode is exponentially damped with
a long characteristic time $\tau_{\rm M} = 2600$s. The first spatial
harmonic at $K_{\rm M}=2 \cdot (2\pi/L_{\rm p})$ can also be extracted:
it decays 4 times faster, as expected from linear dynamics. Higher
harmonics are negligeable.

\begin{figure}
\begin{center} 
\includegraphics[width=7cm]{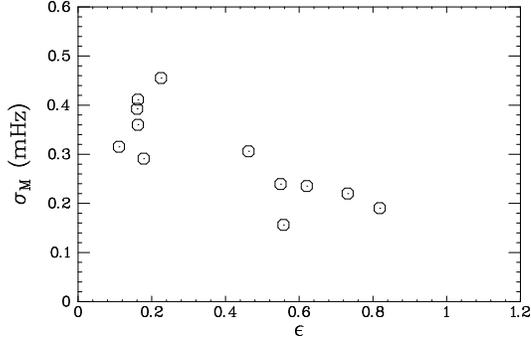} 
\end{center}

\caption{Temporal decay rate of low wavenumber modulations ($K_{\rm
M}=2\pi/L_{\rm p}$) {\it versus} $\epsilon$. Those data are obtained for
patterns of mean wavenumber close to $k_{\rm c}$, {\em i.e.}, within $\pm
2\pi/L_{\rm p}$. For $\epsilon$ close to $\epsilon_{\rm E} \sim 1$,
modulations do not decay anymore and Eckhaus instability takes place.
Measurements close to $\epsilon_{\rm E}$ are very long and difficult to
perform because the pattern can break quite spontaneously. }

\label{fig:sigmaeps}
\end{figure}

Once the modulation has relaxed, and the UHW (uniform hydrothermal wave)
regime is reached, one can produce new modulations very easily, for
example by dropping a fluid droplet in the channel or simply by touching
the free surface with the needle devoted to fluid thickness
measurements. We also frequently observe very small spontaneous
modulations, due to experimental noise, that travel along the cell for
several hundred seconds. When $\epsilon$ is varied from a relaxed UHW
state, another modulated wave is produced: the larger is the $\epsilon$ jump,
the larger the initial modulation amplitude is. Using droplets, very
strong modulations can be initiated: the wavenumber modulation can reach
five percent of the mean wavenumber value \cite{mukchi98}. The very low
damping rate is the signature of the presence of a secondary bifurcation
for slightly higher $\epsilon$: the higher $\epsilon$, the larger
$\tau_{\rm M}$. For different $\epsilon$ we plot the damping rate
$\sigma_{\rm M} = 1/\tau_{\rm M}$ in Fig.~\ref{fig:sigmaeps}. The
modulation wavenumber is $K_{\rm M}=2\pi/L_{\rm p}$ ---the lowest
achievable in a finite periodic box~\cite{tucbar90}--- whatever
$\epsilon$; the modulation frequency $\Omega_{\rm M}$ varies slowly with
$\epsilon$, and this is discussed below (section~\ref{sec:vg_annulus}). We
observe $\sigma_{\rm M}$ to decrease with $\epsilon$ and to approach
zero for $\epsilon = \epsilon_{\rm E} \simeq 1$. At this point, the wave
pattern is unstable with respect to the modulational Eckhaus
instability.

We wish to emphasize that measurements close to $\epsilon_{\rm E}$ are
very long and difficult to perform because the pattern can break quite
spontaneously. This is the reason why the transition is not clearly
visible in Fig.~\ref{fig:sigmaeps}. We will show that the fragility of
the pattern is due to the nature of modulated wave solutions (see
discussion in section \ref{sec:supeck}). It is also obviously related to
the metastability of the UHW with respect to changes of their integer
mean wavenumber $k$: when $\sigma_{\rm M}$ gets smaller, the pattern
becomes less stable and small fluctuations or perturbations due to the
presence of the operator may induce a pattern change. 

The next section describes the modulated waves above $\epsilon_{\rm E}$
wich are even more fragile.

\subsubsection{Stable modulated waves above Eckhaus onset}
\label{sec:mwstable}

\begin{figure}
\begin{center} 
\includegraphics[width=12cm]{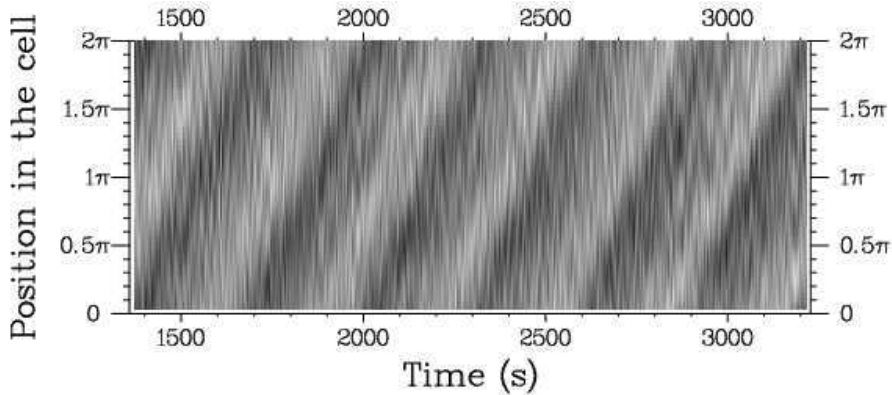} 
\end{center}

\caption{Space-time image of the local wavenumber of the $kL = 55$
hydrothermal wave for $\Delta T = 5.96$~K and $\epsilon = 0.92$. This
wavenumber modulation is non-linearly saturated and propagates steadily
along time. According to Eq.~\ref{eq:defq} we measure: $q = (0.29 \pm
0.05) 2\pi/L_{\rm p}$. Spatial frequencies smaller than the third of the
cell size are filtered.}

\label{fig:imsatur}
\end{figure}

Once $\epsilon$ is set greater than $\epsilon_{\rm E}$ we observe that
the wavenumber modulation at $K_{\rm M} = 2\pi/L_{\rm p}$ grows, then
saturates and persists for several hours (Fig.~\ref{fig:imsatur}). We
have seen no evidence of hysteresis in this transition. It's a
supercritical Hopf bifurcation \cite{mukchi98}. The bifurcated
modulation amplitude is quite small compared to the modulation obtained
either by changing the control parameter, either by forcing. Its
amplitude is typically 5 times the noise level for the phase gradients
in the best case for our standard experimental conditions at $h=1.7$~mm:
the modulation can be detected on the phase gradients, but is too small
to be measurable on the carrier wave amplitude. The amplitude of the
wavenumber modulation represents typically $0.5\%$ of the mean
wavenumber value. 

Those modulated waves have been plotted on the stability diagram
(Fig.~\ref{fig:diagstab1}). We notice that these stable modulated waves
around $\epsilon_{\rm E}$ do not represent a well defined region in the
$(\epsilon,k)$ diagram. They are observed in a very narrow band in
$\epsilon$, the thickness of which is comparable to the noise level in
this region. This noise level is very high due to the extreme
sensitivity of modulated waves to noise and perturbations, which makes
it very hard to reproduce exactly the experimental conditions. So, when
the cell contains a supercritically saturated modulated wave pattern,
the control parameter has to be changed by very small steps. Otherwise,
the resulting perturbations produce strong modulations, break the
pattern, and change its mean wavenumber. In this region of parameters,
it is also impossible to measure the fluid thickness without perturbing
and then breaking the pattern. We estimate the band of stable modulations to
extend from $\epsilon_{\rm E}$ to $\epsilon_{\rm SN}$, $\epsilon_{\rm
E}$ being typically close to unity around $k_{\rm c}$ and slowly
decreasing with $k$, and $(\epsilon_{\rm E}-\epsilon_{\rm SN})$ is
typically $0.2$ or $0.3$. The upper bound $\epsilon_{\rm SN}$ is named
by reference to the saddle-node bifurcation of MAWs
\cite{brutor01a,brutor01b} a point which will be emphasized in the
discussion below (section~\ref{sec:annulus_disc}).

Fragility seems thus to be the principal feature of those modulated
patterns at $h=1.7$~mm and $k \sim k_{\rm c}$. Whereas plane- wave
patterns may be strongly perturbed near the waves onset ---$\epsilon
\sim 0$, see Fig.~\ref{fig:mwdamped}---, the fragility to perturbations
increases when $\epsilon_{\rm E}$ is approached, and becomes extreme in
the supercritical Eckhaus band. The next paragraph describes how these
modulated wave patterns spontaneously die above $\epsilon_{\rm SN}$.

\subsubsection{Exploding modulated waves and spatio-temporal dislocations}

\begin{figure}
\begin{center} 
\includegraphics[width=15cm]{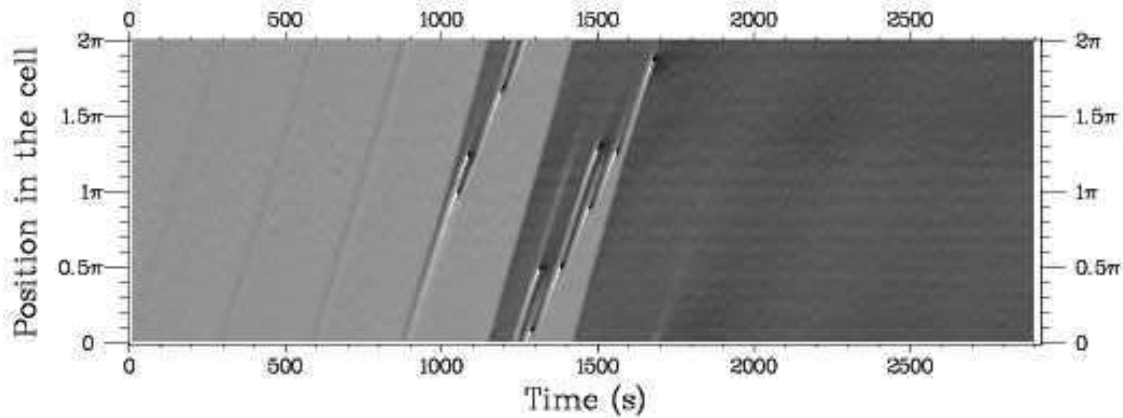} 
\end{center}

\caption{Transient evolution of a $k = 55(2\pi/L_{\rm p})$ Eckhaus
modulated wave pattern toward a stable $k = 45(2\pi/L_{\rm p})$ UHW
pattern, when $\epsilon$ grows from $0.92$ (pattern on
Fig.~\ref{fig:imsatur}) to $1.21$ above $\epsilon_{\rm SN}$. This image
represents the local wavenumber without filtering or enhancement. At the
first stage, the $K_M = 2\pi/L_{\rm p}$ modulation travels at constant
velocity. Its maximum peak value increases slowly (characteristic growth
time $700$~s) while the amplitude of the fundamental ($2\pi/L_{\rm p}$)
remains constant. Then, after $t=870$~s, both max amplitude and
fundamental amplitude start growing on a faster scale ($\tau_{\rm
M}=100$~s). Then, higher spatial frequencies start growing and generate
ten successive traveling wavenumber peaks. The wavenumber peaks
correspond to growing phase jumps: once they reach $\pi$, the local
amplitude (not shown) simultaneously attaining zero and the local
wavenumber changing sign, they create ten space-time defects which each
annihilate one wavelength of the wave-pattern. The resulting $k =
45(2\pi/L_{\rm p})$ pattern is Eckhaus stable: the remaining traces of
modulations decay.}

\label{fig:holes}
\end{figure}

By increasing the constraint above this narrow stability band $(\epsilon
> \epsilon_{\rm SN})$, we recover the usual Eckhaus behavior,
illustrated in Fig.~\ref{fig:holes} which represents the evolution of a
$k = 55(2\pi/L_{\rm p})$ modulated carrier pattern when $\epsilon$ is
increased above $\epsilon_{\rm SN}$. We observe the spontaneous growth
of high wavenumber modulations leading to space-time defects or
amplitude holes \cite{legjan92} where the amplitude goes to zero and the
phase jumps by $\pi$ (see Fig.~\ref{fig:holes} and caption for details).
After ten phase jumps, the mean wavenumber of the carrier TW has
decreased to $k = 45(2\pi/L_{\rm p})$ and the pattern relaxes towards a
non-modulated UHW state. Note that this example corresponds to a strong
increase of the control parameter $\epsilon$. For smaller steps above
$\epsilon_{\rm SN}$, the transient state lasts much longer: many more
traveling modulations are observed, whereas only five or six of them
reach the zero-amplitude level and change the carrier pattern
wavenumber. This recalls the slow chaotic state \cite{kol92} for Eckhaus
transition of subcritical TW. The non-saturated growth of
waves-modulations have been described as the basic mechanism for the
development of the Eckhaus instability in several systems
\cite{kol92,baxeat92,janpum92,liueck99,lewpro95} leading to
spatio-temporal defects and subsequent variation of the mean wavenumber.
The only difference in our system is the basic state which is already
slightly modulated. This is quite invisible on the spatio-temporal
diagrams (Fig.~\ref{fig:holes}): the original traveling modulation is
tiny and would need a strong contrast enhancement to appear on the
picture. Notice also the wavenumber of the growing modulations which
rapidly increases to typically $K_{\rm M} \sim 6$ to $8 (2\pi/L_{\rm
p})$ instead of $(2\pi/L_{\rm p})$. A very interesting observation has
been made by Liu and Ecke \cite{liueck99} in rotating convection: the
further the control parameter is increased into the unstable Eckhaus
band, the more the pattern number of wavelengths is changed. Our system
currently looses as much as six to ten wavelengths. This phenomenon
is probably favored by the existence of the supercritical band:
when defects appear, the distance to $\epsilon_{\rm E}$ is already
finite. This effect depends probably also of the slopes of
$\epsilon_{\rm E}(k)$ and $\epsilon_{\rm SN}(k)$.

\begin{figure}
\begin{center} 
\includegraphics[width=5.5cm]{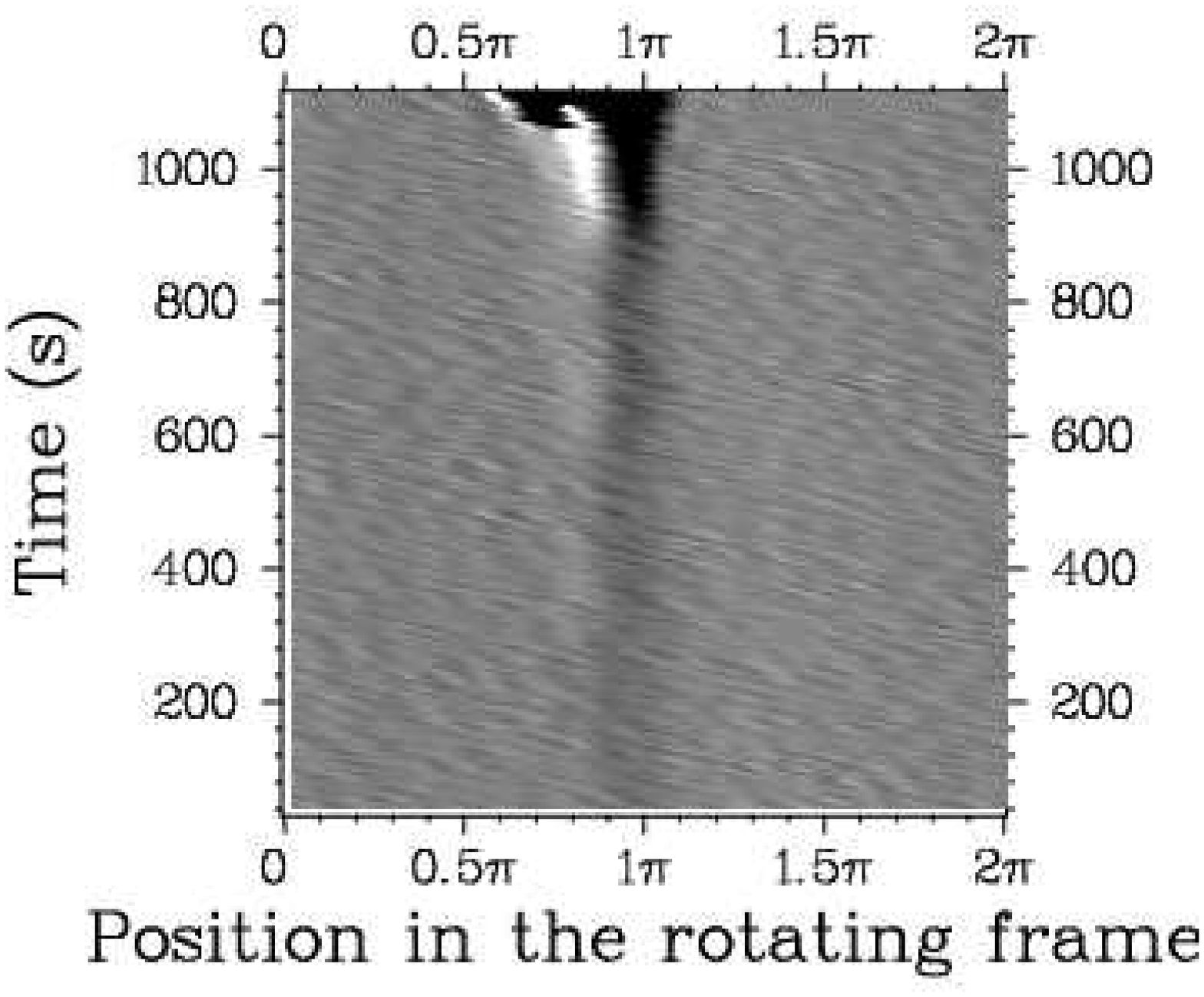}%
\includegraphics[width=8cm]{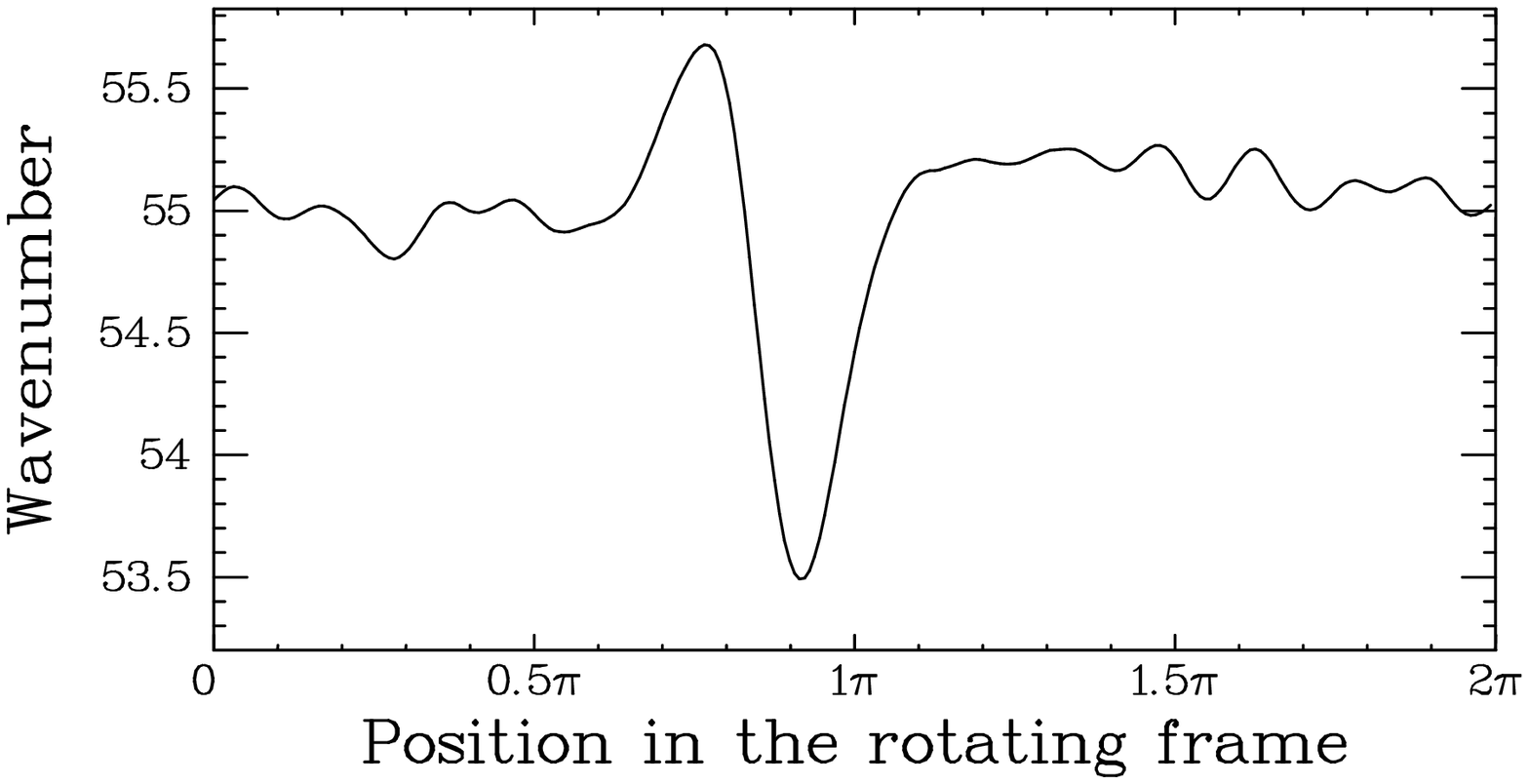} 
\end{center}

\caption{Same wavenumber data as in Fig.~\ref{fig:holes}, presented in
the frame rotating at the modulation velocity. From the spatio-temporal
image, an averaged profile ($600\,{\rm s} < t < 800\,{\rm s}$) of the
modulation phase gradient is extracted.}

\label{fig:holes_rot}
\end{figure}

Fig.~\ref{fig:holes_rot} presents the growing modulation in the rotating
frame where it is stationary. The shape of the profile does not vary
very much until the first defect appears. Such profile is very similar
to the typical phase-gradient profiles of MAWs \cite{brutor01a}.

Once the pattern has relaxed to a lower wavenumber, we may again
increase $\epsilon$. Eckhaus modulations and pattern breaking occurs
again, until the mean wavenumber reaches $44$ or $45(2\pi/L_{\rm p})$.
However, below $49$ or $50(2\pi/L_{\rm p})$, the nature of the
modulations changes dramatically as will be shown in
section~\ref{sec:square}.

\subsubsection{Velocity of modulated waves and group velocity}
\label{sec:vg_annulus}

\begin{figure}
\begin{center} 
\includegraphics[height=5cm]{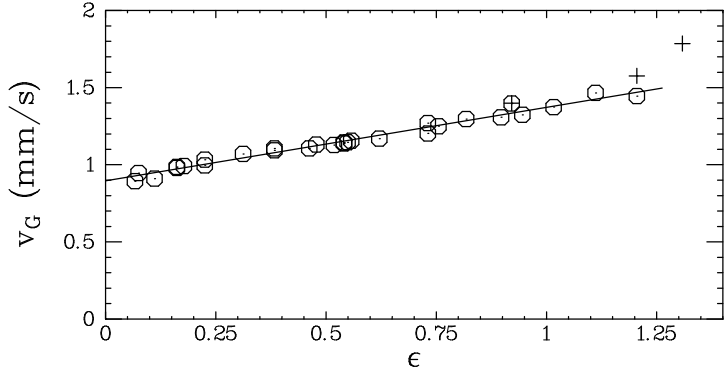} 
\end{center}

\caption{Velocity $v_{\rm G}=\Omega_{\rm M}/K_{\rm M}$ of small
wavenumber ($K_{\rm M}=2\pi/L$) modulations {\it vs.} $\epsilon$.
Circles ($\circ$) correspond to modulations decaying towards UHW.
Plusses ($+$) correspond to stable or metastable (far before the
apparition of space-time defects) modulations. These data are taken in
the central wavenumber-band around $k_{\rm c}$, {\em i.e.}, for $53 \leq
kL/2\pi \leq 57$. }

\label{fig:vg_annulus}
\end{figure}

The observation of traveling modulations of waves gives a basic
information: the phase velocity of the modulation. As far as small
smooth modulations can just be considered as perturbations of the wave
envelope, their velocity is equivalent to the hydrothermal wave group
velocity. This is the case for damped modulated waves below
$\epsilon_{\rm E}$, at least just before they vanish. In fact the
velocity is almost independent of the amplitude of the modulation. We
extract this velocity from the study of the modulation frequency
$\Omega_{\rm M}$, for wavenumbers in the very central band around
$k_{\rm c}$, {\em i.e.}, for $53 \leq kL/2\pi \leq 57$. The group velocity
$v_{\rm G}=\Omega_{\rm M}/K_{\rm M}$ is plotted on
Fig.~\ref{fig:vg_annulus}. Extrapolation at the wave onset is used to
measure the value of $s$ given in Eq.~\ref{eq:coeff-s}. The variation of
$\Omega_{\rm M}$ (or $v_{\rm G}$) with $\epsilon$ may also be used in the
fit of the HOCGL coefficients, because it is one of the terms in the
development of the eigenvalue $\sigma$ for the secondary instability mode:

\begin{equation}
\sigma = -D_{\parallel}K_{\rm M}^2 + i v_{\rm G}K_{\rm M} + {\rm O} (K_{\rm M}^3)
\label{eq:sigma}
\end{equation}

where $D_{\parallel}$ is the phase diffusion coefficient for modulational
perturbations, directly related to the damping of the modulation
amplitude (Fig.~\ref{fig:sigmaeps}).

Fig.~\ref{fig:vg_annulus} also presents a few data concerning supercritical
traveling modulations. Some values of $v_{\rm G}$ are a bit larger
than what would be extrapolated from the damped modulations data. This
may be the sign of a particular selection of the modulation velocity
\cite{brutor01a,brutor01b} (see discussion in section~\ref{sec:annulus_disc}).

\begin{figure}
\begin{center} 
\includegraphics[height=4.5cm]{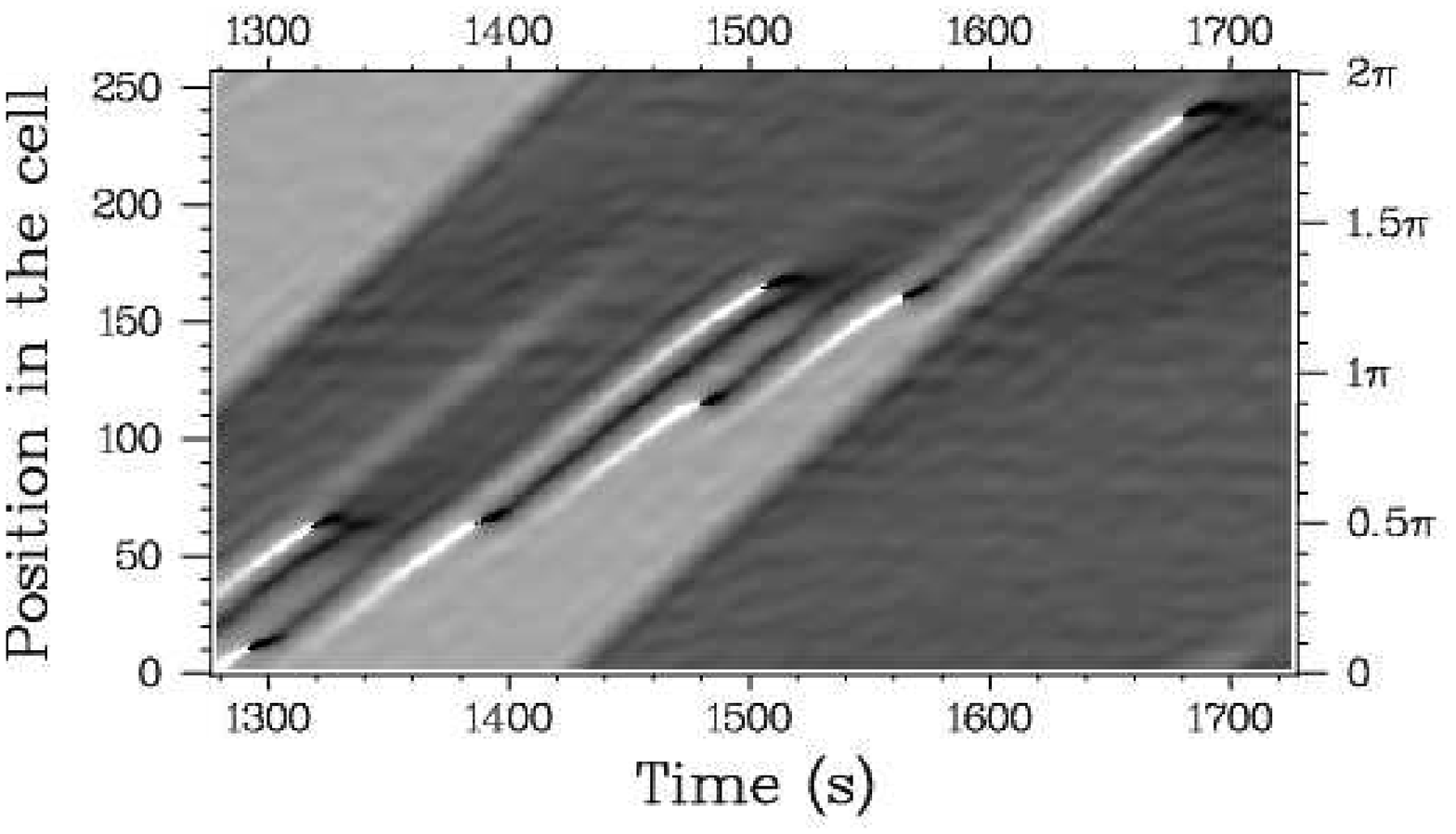}%
\includegraphics[height=4.5cm]{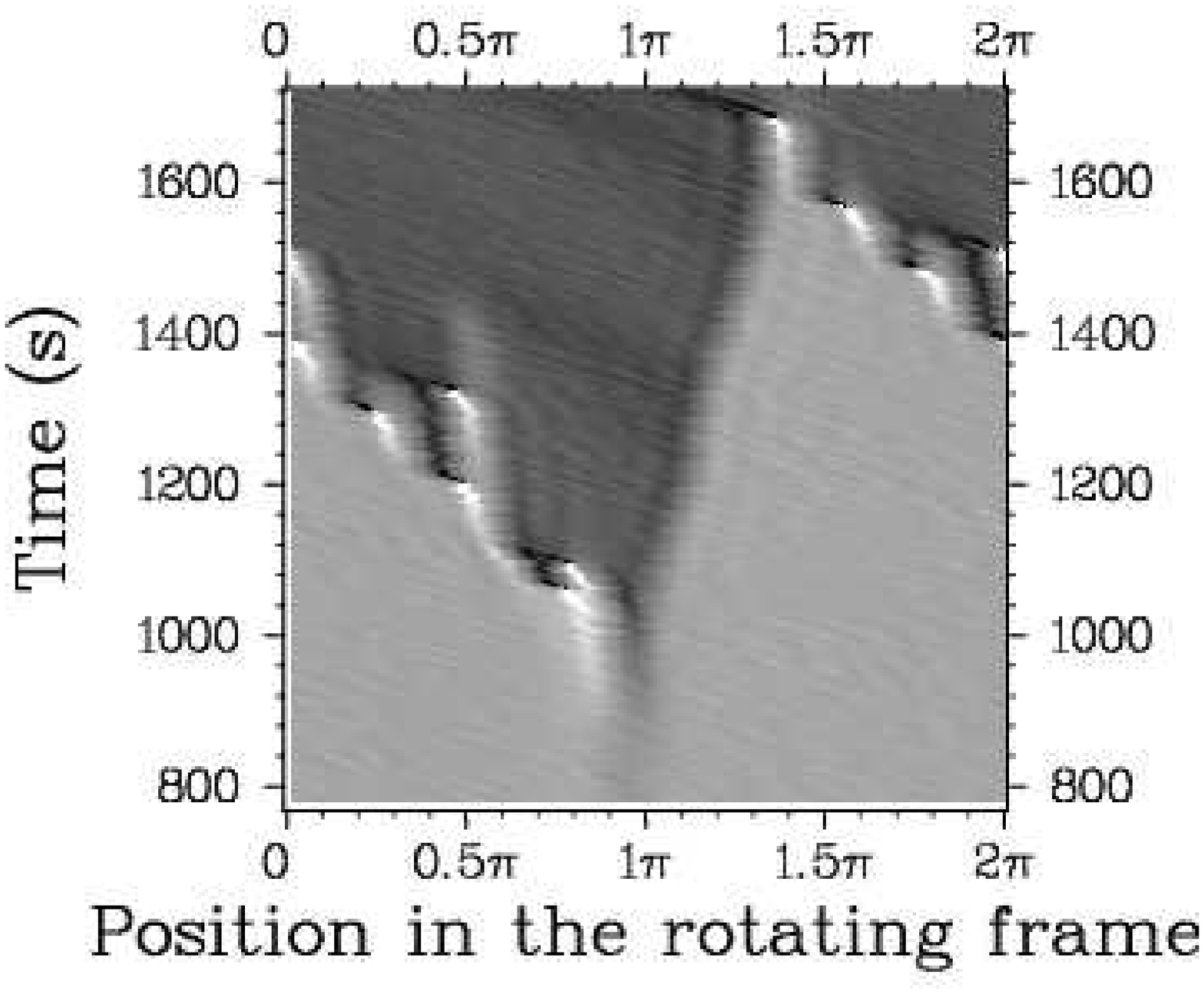} 
\end{center}

\caption{Same wavenumber data as in Fig.~\ref{fig:holes}.
(a) Time detail of the pattern around defect cores in the laboratory frame: 
please note how the hole cores slow down before disappearing.
(b) Global view presented in a frame rotating at the initial
modulation velocity. Using much lower contrast than in
Fig.~\ref{fig:holes_rot}, one can follow the defects during their whole
life: the hole core velocity is negative in this frame.
}

\label{fig:holes_zoom}
\end{figure}

{\it A contrario}, above $\epsilon_{\rm SN}$, we observe the modulation
velocity to decrease at the end of the growth phase, just before the
nucleation of spatio-temporal dislocations (Fig.~\ref{fig:holes}): a
smooth trace of each modulation keeps traveling at the group velocity
until it disappears, while the sharp phase-gradient peaks slow down, in
the laboratory frame, near the defects core (Fig.~\ref{fig:holes_zoom}).

\subsection{Large square modulated waves far from $k_{\rm c}$ }
\label{sec:square}

\begin{figure}
\begin{center} 
\includegraphics[width=14cm]{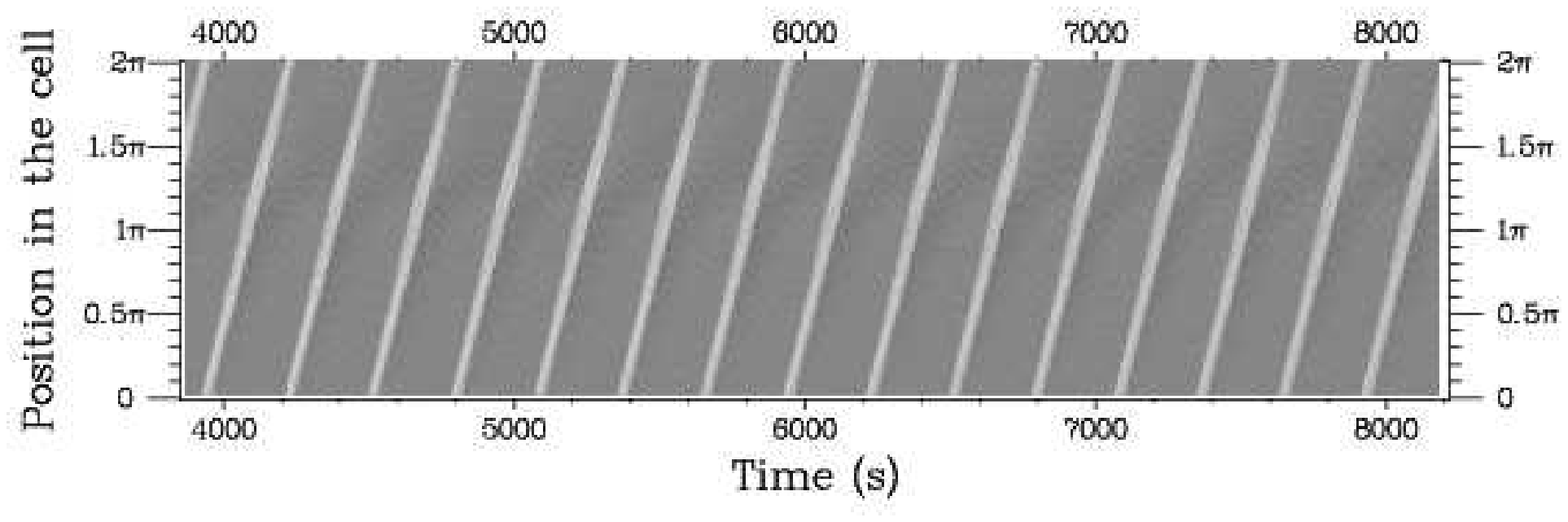}
\includegraphics[width=10cm]{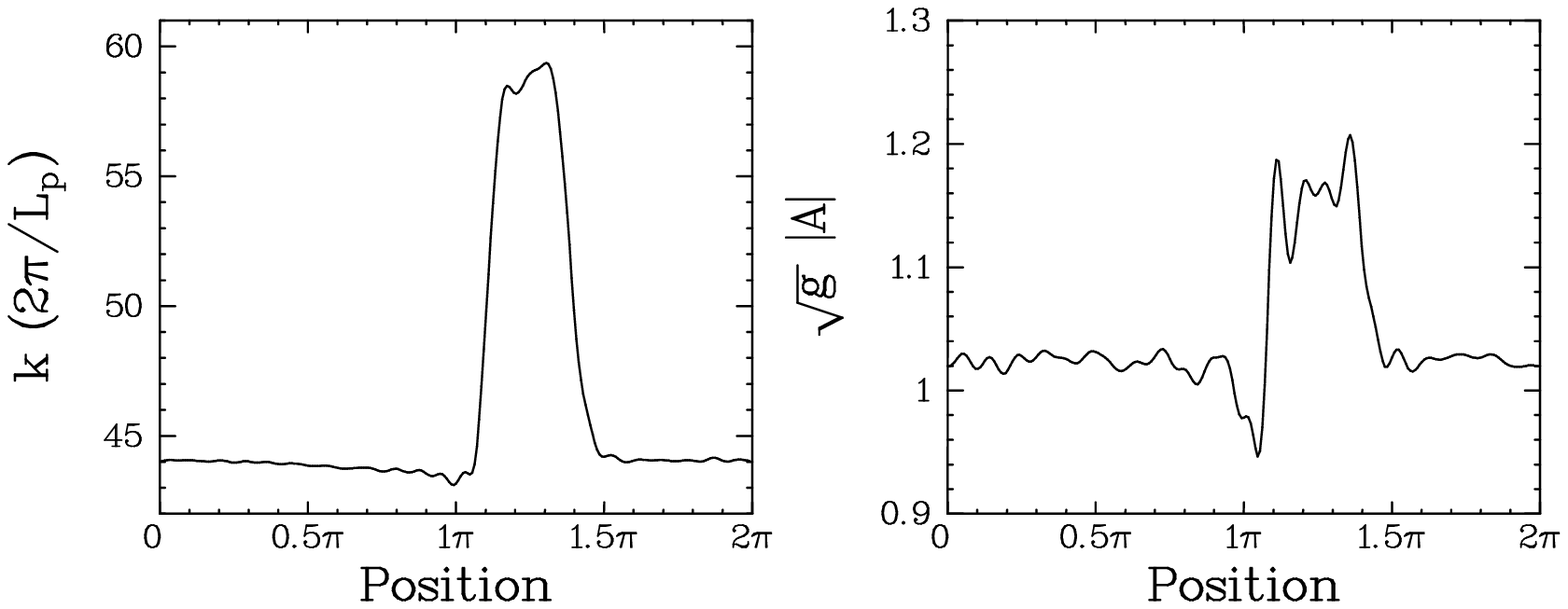} 
\end{center}

\caption{Strong square modulations for $\epsilon=1.32$ ($\Delta T=
7.2$~K) and mean wavenumber $k=46(2\pi/L_{\rm p})$. The local and
instantaneous wavenumber is displayed on the space-time diagram (a).
This pattern may be compared to two imbedded regions of quasi-uniform
wavenumber $k=44(2\pi/L_{\rm p})$ and $k=59(2\pi/L_{\rm p})$ traveling
together. The spatial wavenumber profile (b) and wave-amplitude profile
(c) are averaged in the rotating frame over eight periods. }

\label{fig:square}
\end{figure}

Traveling wave patterns at low wavenumber $k \lesssim 45(2\pi/L_{\rm
p})$, below the central band are also unstable with respect to Eckhaus
instability, but in somewhat different conditions. Once the Eckhaus
onset is crossed, very strong square modulations appear. The name {\em
square} is chosen because the local wavenumber and frequency signals
along space and time represent a square signal, or at least exhibit a
sharp front. The amplitude of these signals saturates at a very high
value compared to the previous supercritical case: once the modulation
passes at a point, the local phase-gradient changes of typically 10 to
20 percents of its mean value (compare to $0.5\%$ in the previous case
close to $k_{\rm c}$). The modulations appear spontaneously with a
large amplitude: it is probably a subcritical bifurcation with a high
order non-linear saturation, {\em i.e.}, again a very different pattern
from the classical subcritical Eckhaus transition~\cite{kol92}!
Another feature of the square modulations is their spatial wavenumber
$K_{\rm M}$ or spatial period $P=2\pi/K_{\rm M}$. We observed $P$ to
vary from $L_{\rm p}$ to $L_{\rm p}/5$, instead of being uniformly
$L_{\rm p}$ in the supercritical case. Fig.~\ref{fig:square}a present
such $P=L_{\rm p}$ square pattern. The wavenumber profile and the
amplitude profile along the channel are shown in Fig.~\ref{fig:square}b
and Fig.~\ref{fig:square}c respectively.

Those square patterns are not systematically observed when the
experiment is reproduced. Moreover, they may decay very slowly
(typically over a day) and vanish. No systematic study has been realized
over such time scale. If the control parameter is increased
sufficiently, the sharp high wavenumber part of the signal generates
spatio-temporal dislocations and the pattern looses $L_{\rm p}/P$
wavelengths. The nature and stability of square modulated waves is an
open question.

\subsection{Spatio-temporal chaos at high $\epsilon$ and far from 
            $k_{\rm c}$: toward a globally restored symmetry}
\label{sec:chaos}

\begin{figure}
\begin{center} 
\includegraphics[width=7cm]{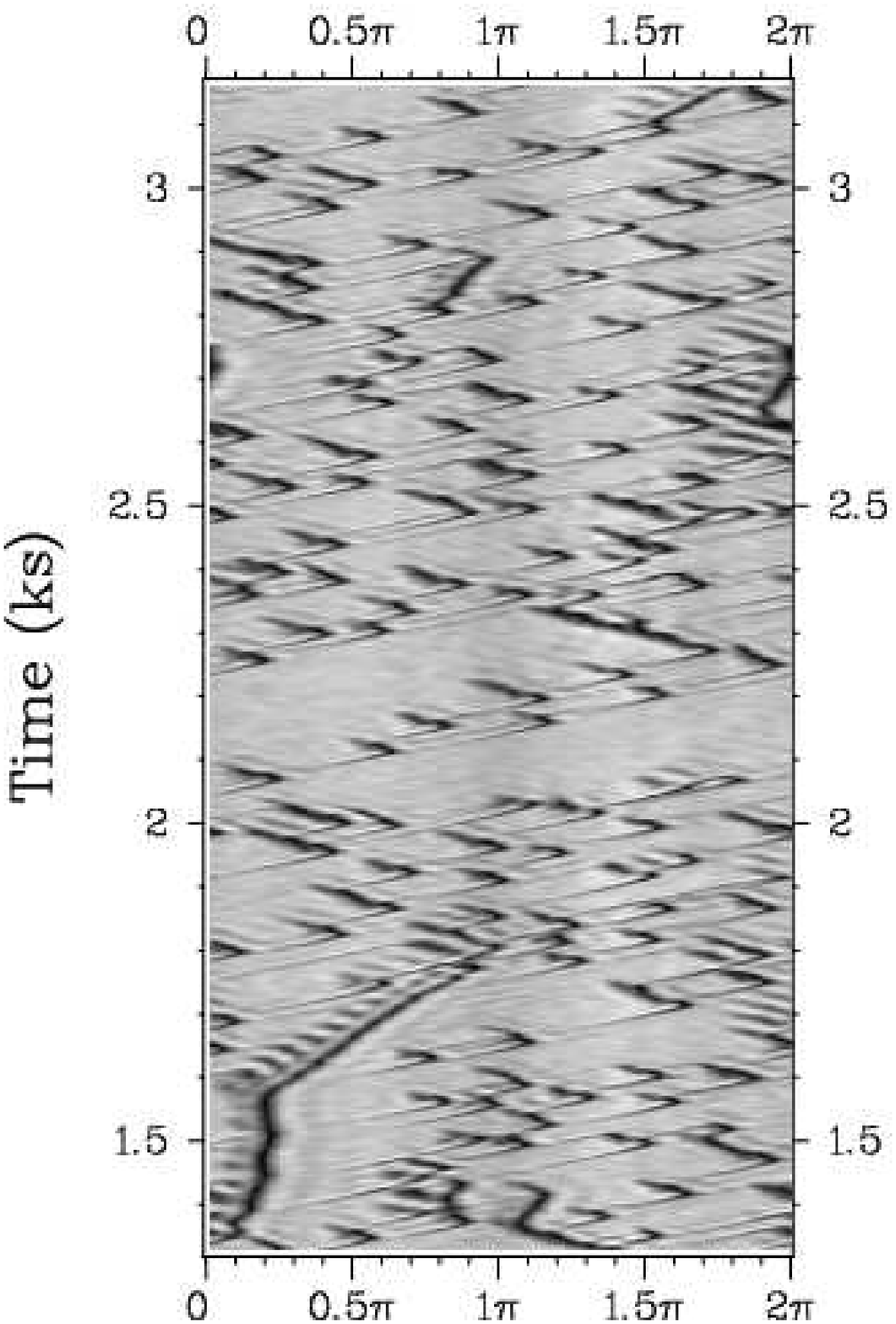}%
\includegraphics[width=7cm]{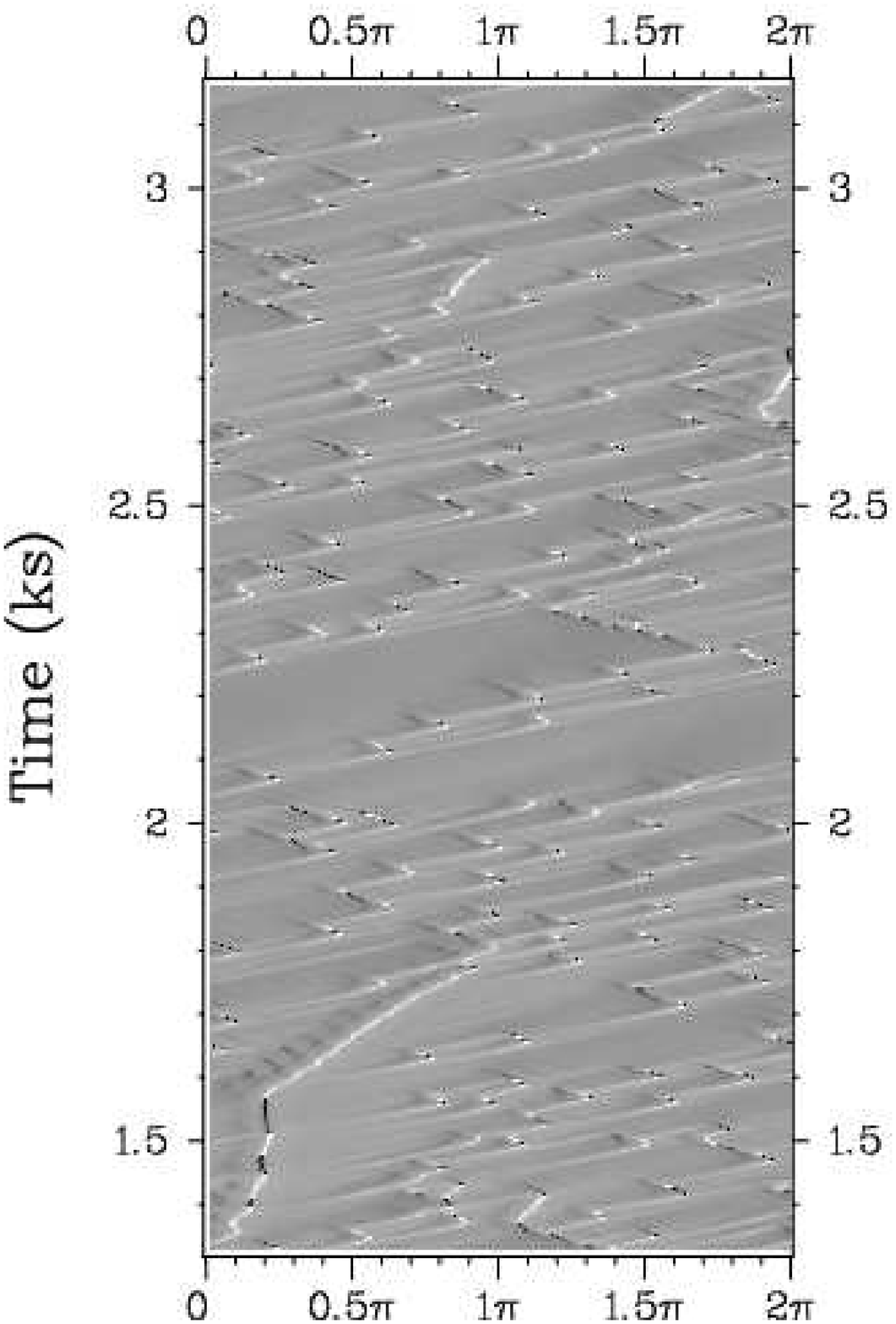} 
\includegraphics[width=7cm]{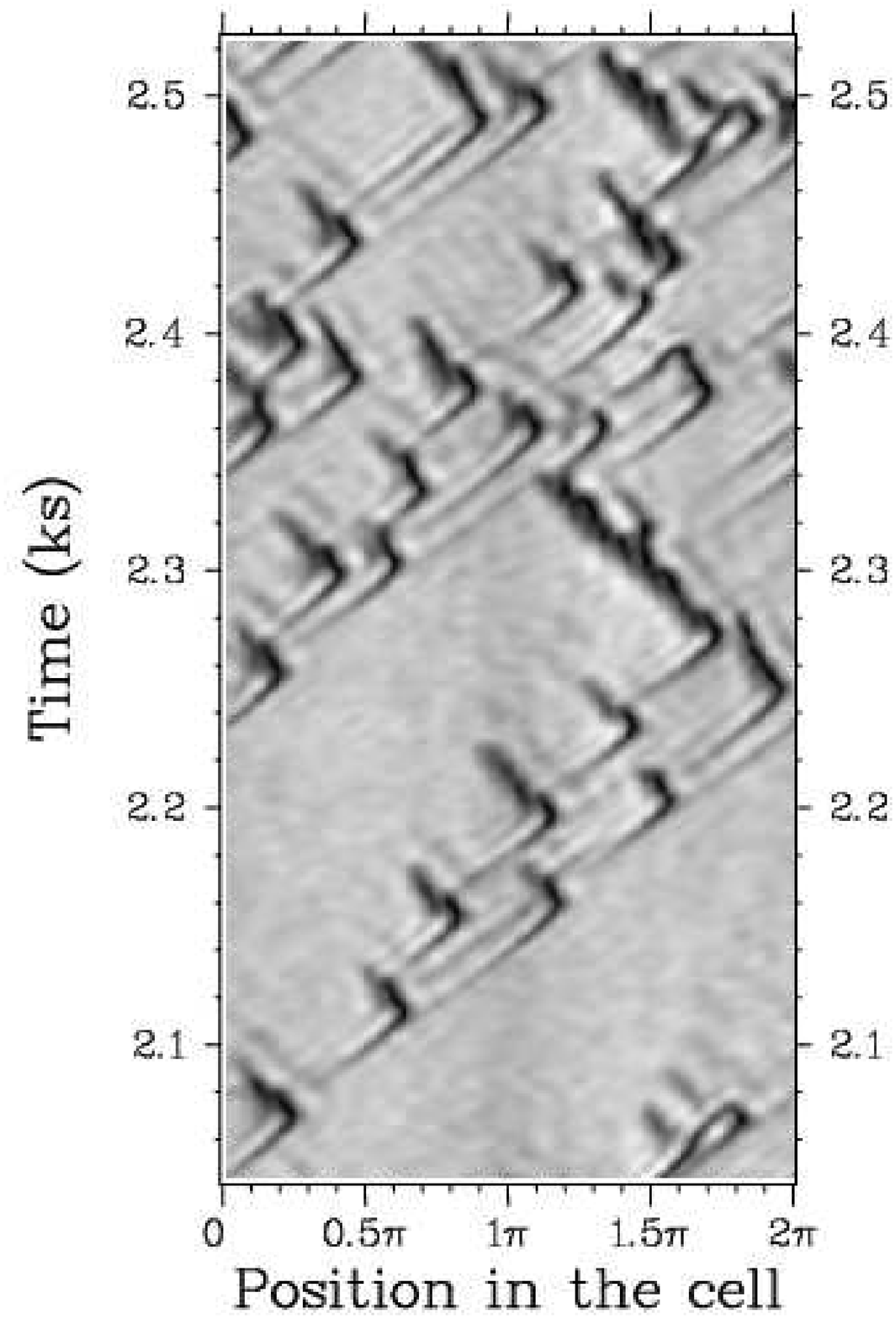}%
\includegraphics[width=7cm]{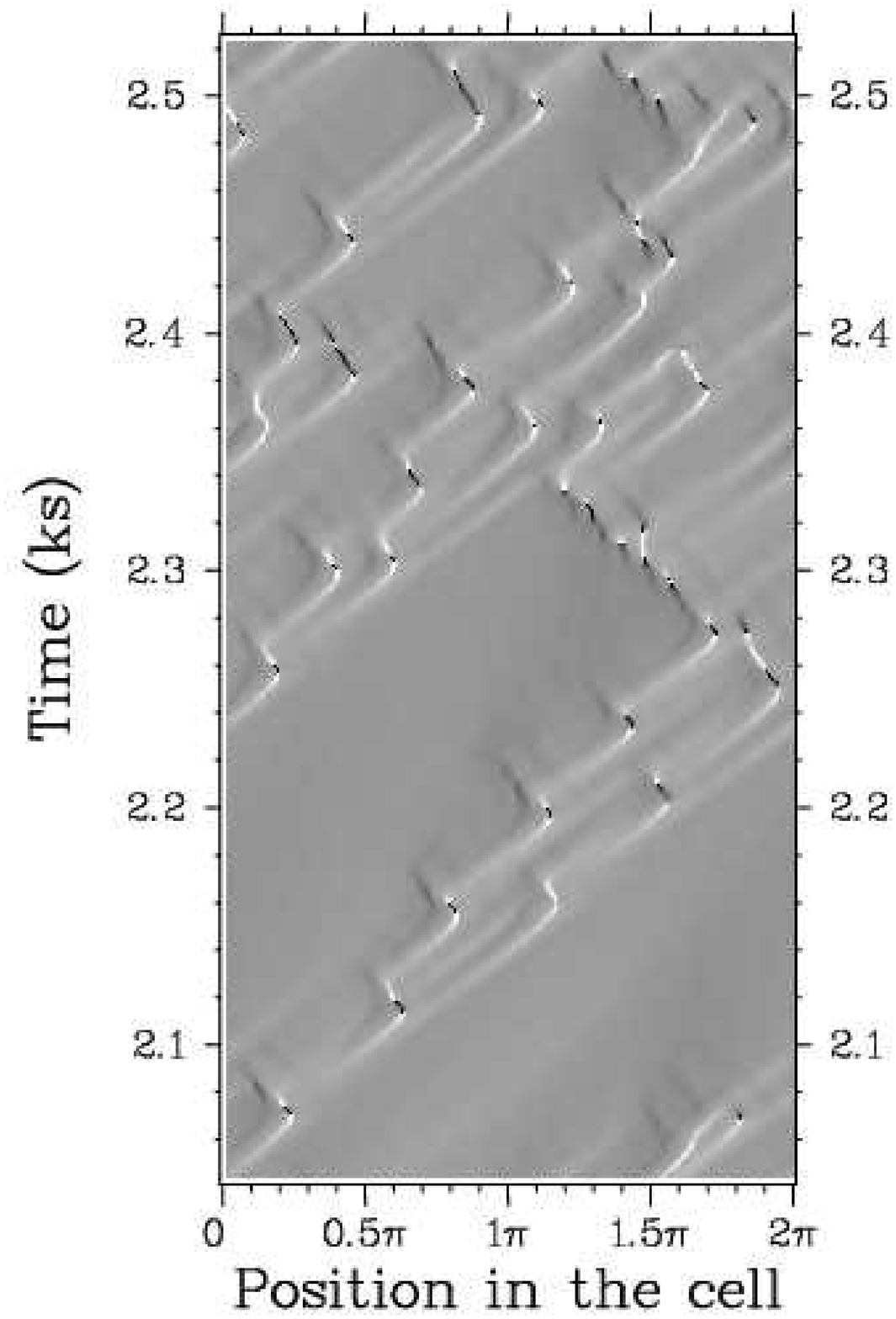} 
\end{center}

\caption{Spatio-temporal chaos due to small scale traveling holes and
modulations. Spatio-temporal diagrams of the local and instantaneous
amplitude $|A|$ (left) and of the local wavenumber (right) of the
right-traveling wave for $\epsilon=3.11$. The gray scale for amplitude
data is proportional to $|A|$: zero amplitude appears black. The mean
wavenumber fluctuates around $k=43(2\pi/L_{\rm p})$. A general view is
given (top). A region showing the most common holes patterns is enlarged
(bottom). Some slower patterns on the general view correspond to couples
of source and sink, traveling very close to each other and remaining
parallel. Those objects cannot be studied without the left-traveling
wave information. These diagrams are presented in the laboratory frame.
We estimate the group velocity as being the velocity of small amplitude
modulations far from the $A=0$ defects cores. }

\label{fig:turbholes}
\end{figure}

In the above sections, description of states have been made which
---except for exploding modulations above $\epsilon_{\rm SN}$---
presents only phase dynamics. As a consequence, the integer mean
wavenumber $k$ was considered as a constant parameter in the dynamics.
In this section, we will describe amplitude-chaotic patterns which are
not concerned anymore by this constraint: the mean wavenumber may
fluctuate since the presence of topological defects allow the phase to
change by steps. This is an example of defect-chaos \cite{shrpum92}.

Once the wavenumber decreases below $43$ or $44(2\pi/L_{\rm p})$, no
periodically modulated waves are ever observed. The $k=43(2\pi/L_{\rm
p})$ UHW can be driven up to $\epsilon \sim 2.6$, {\em i.e.}, very far
from the onset of waves. For higher $\epsilon$ the system transits to
disordered patterns instead of periodically modulated waves; then, the
mean wavenumber of the pattern ceases to behave as a constant of the
dynamics. Three different regimes may be encountered. They have been
studied in the range $2.8 \lesssim \epsilon \lesssim 3.9$. Those
regimes need very long observations in order to be characterized: a
very slow temporal intermittency regime is observed which makes those
chaotic states appear and disappear, intercalated with metastable UHW
in a very-low-wavenumber range ($39 \lesssim kL_{\rm p}/2\pi \lesssim
43)$. The characteristic time of this intermittency is of order
of 2-10 hours.

The first regime shows UHW being densely invaded by traveling holes
(Fig.~\ref{fig:turbholes}). These objects behave as localized modulated
waves existing over much smaller scales than the low wavenumber
modulations described above. They present a complex dynamics and
interact together. They travel in the same direction as the carrier
wave, but once they reach their minimal amplitude, very close to zero,
the direction of propagation may reverse. Traces of the
counter-propagating wave are sometimes observed in this backward
traveling part of the hole. Mean wavenumbers are typically around $44
(2\pi/L_{\rm p})$. 

\begin{figure}
\begin{center} 
\includegraphics[height=10cm]{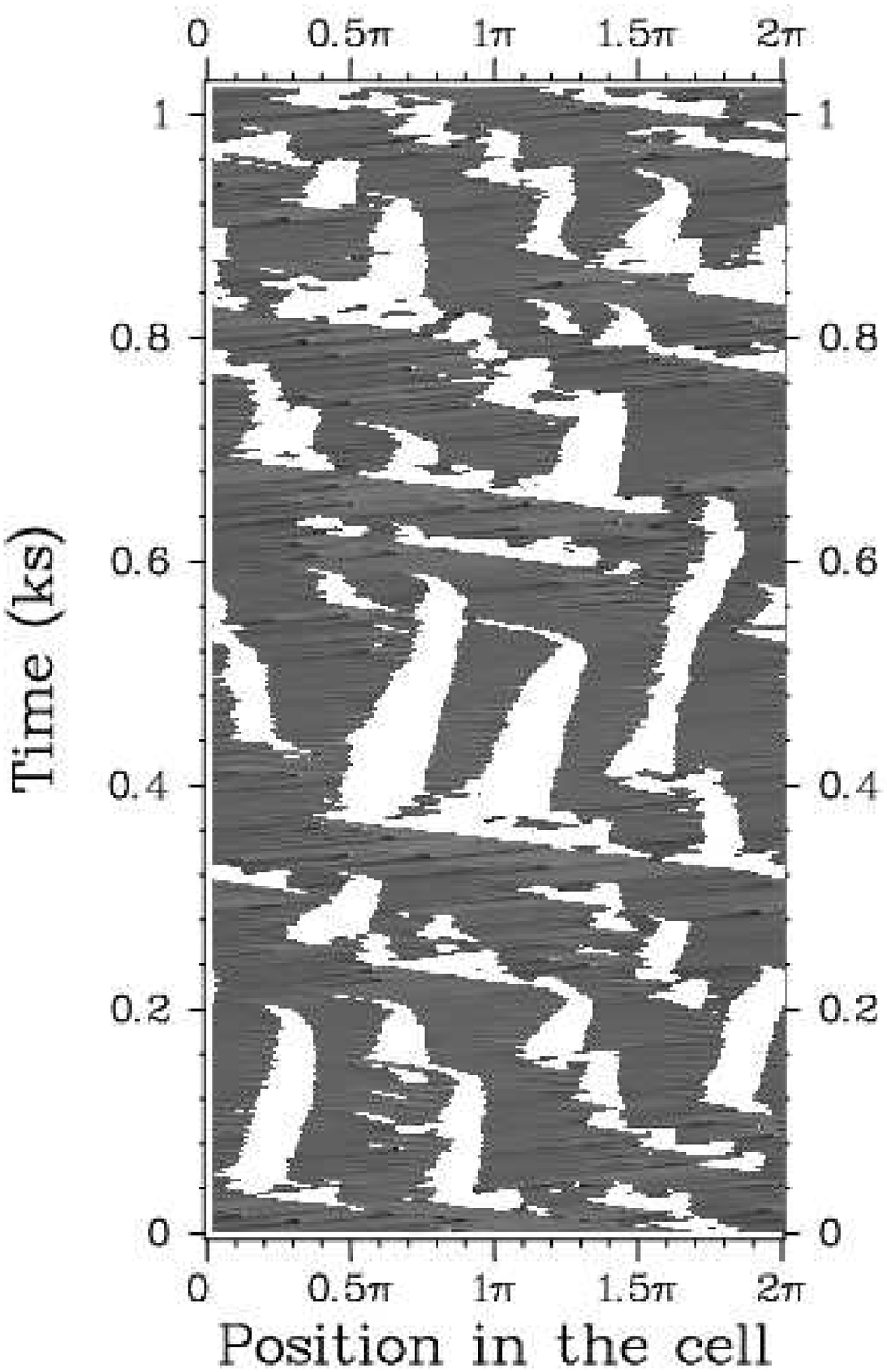}%
\includegraphics[height=10cm]{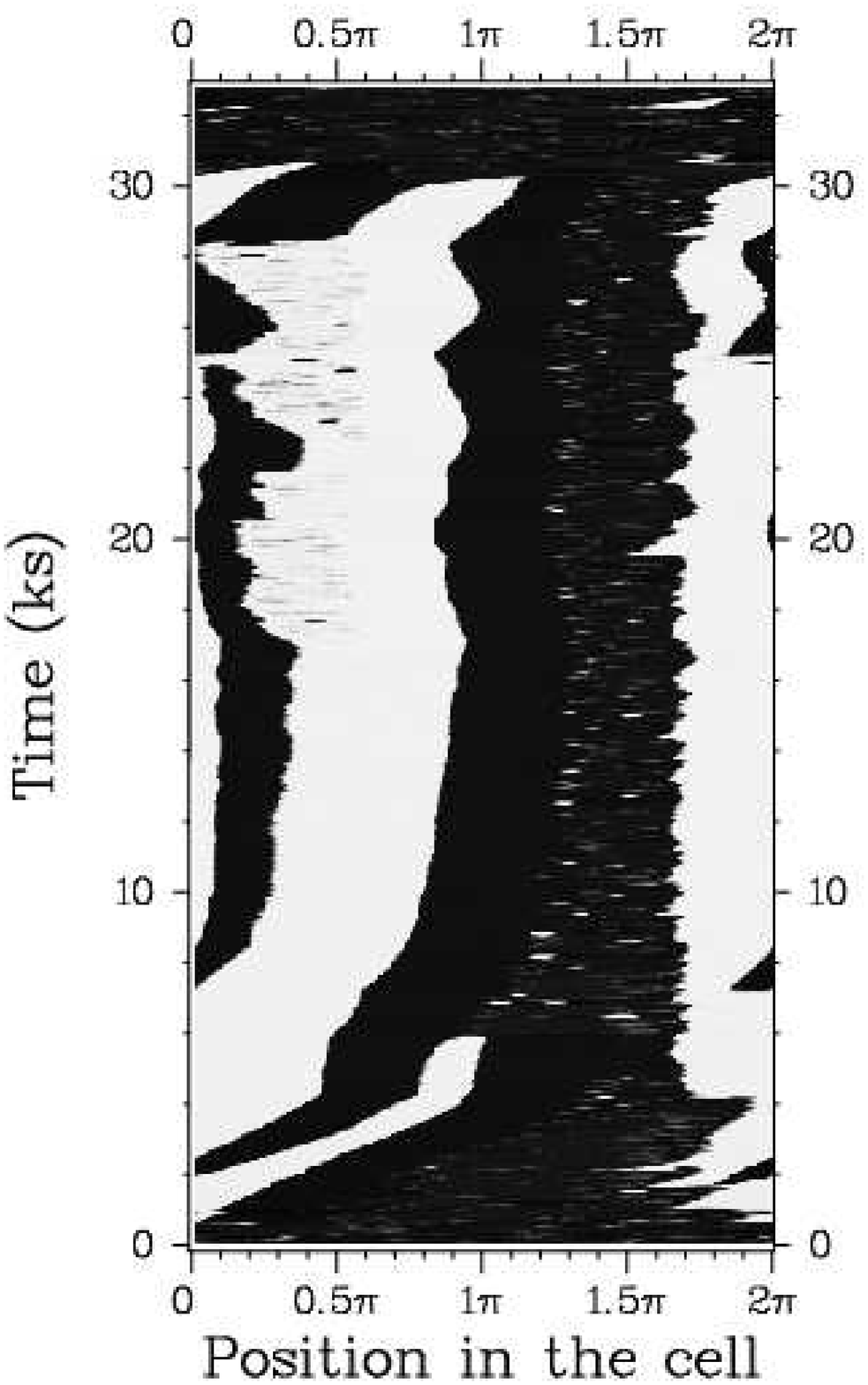} 
\end{center}

\caption{Spatio-temporal chaos due to sources and sinks dynamics, {\em i.e.},
counter-propagating waves competition. This figure shows spatio-temporal
diagrams of the local and instantaneous frequency of the wave-patterns
after a Hilbert transform in space. Right (resp. left) traveling waves
appears as dark (resp. light) areas. The left image ($\epsilon=3.25$ and
$k \sim 43(2\pi/L_{\rm p})$) presents a dominant right-traveling wave
with bubbles of left-traveling wave in a $3/4$--$1/4$ mean ratio. This
ratio reverses several time during the complete data acquisition. The
contrast is adjusted to show traveling modulations and holes in the dark
right-traveling wave region, and white left-traveling regions are
saturated. The right image ($\epsilon=2.84$ and $k \sim 46(2\pi/L_{\rm
p})$) presents a symmetric pattern: right- and left-traveling wave in a
$1/2$-$1/2$ mean ratio. Please note the length of the presented data set
(9 hours) which reveals the temporal intermittency: the right-traveling
wave dominates at the beginning and at the end. }

\label{fig:turbPS1/4}
\label{fig:turbPS1/2}
\end{figure}

The second regime contains right- and left-propagating waves domains in
typically $1/4-3/4$ proportion (Fig.~\ref{fig:turbPS1/4}, left). The $x
\mapsto -x$ symmetry, broken at the hydrothermal wave threshold is thus
partly restored at a global level. The minor wave appears in small
``bubbles'' separated from the dominant waves by traveling sources and
sinks. One can note that, during the main lifetime of the
minor-wave-regions, the source velocity is selected while the sink
dynamics is much more erratic until the source/sink pair collides and
annihilates. Such sources/sinks wave patterns show a mean wavenumber
just a bit larger (Fig.~\ref{fig:diagstab1}) than the one of traveling
hole patterns presented above (Fig.~\ref{fig:turbholes}). In fact,
traveling holes and modulations still exist between sources and sinks.
The source velocity is typically one order of magnitude smaller than the
group velocity. Such sources/sinks dynamics looks very similar to the
one reported in numerical simulations for small coupling coefficients by
van Hecke, Storm and van Saarloos (Fig.~10 of Ref.~\cite{hecsto99}) and
by Riecke and Kramer (Fig.~11 of Ref.~\cite{riekra00}).

In the third situation (Fig.~\ref{fig:turbPS1/2}, right), the $x \mapsto
-x$ symmetry appears to be totally restored at the global level: right-
and left-propagating waves regions are equally represented. Thus we
observe the sources to remain fixed along time, being only slightly
affected by small erratic movements of the sinks. Observed patterns
typically exhibit two right-traveling and two left-traveling domains
along the cell (Fig.~\ref{fig:turbPS1/2}, right). The study of the
source velocity selection with right- and left-traveling waves
repartition is under progress and will be published elsewhere. When the
sources and sinks remain at fixed locations, the annular system can be
seen as equivalent to adjacent finite systems \cite{hecsto99} of smaller
sizes. The waves emitted by the sources are modulated waves, very
similar to those reported in the finite box
(section~\ref{sec:rect_seuil2} of II). Those modulations travel across
the whole cell and even pass through the sinks ---which are extended in
space because $|A|$ and $|B|$ vary smoothly in their core. A complex
dynamics can be observed.

Finally, we wish to emphasize the global restoring of the $x \mapsto -x$
symmetry in both regimes of mixed right- and left-traveling waves: while
the wave-proportion fluctuates around its mean value, we often witness
dominant wave direction reversal. Reversal time-scales are comparable to
the temporal-intermittency time-scales described above. At such high
control parameter, the fluctuation level is high and it is thus easy for
the system to transit between symmetric states, as well as between
different dynamical regimes.

\section{Modulated traveling waves: high modulation amplitude solutions in thin layers}
\label{sec:annulus_bis}

Some experiments have been performed in the annular channel with a
smaller fluid depth, down to $0.8$~mm. The fluid height $h$ is the main
length scale of the problem~\cite{garnier00}. The wavelengths, $k_{\rm
c}^{-1}$ and $\xi_0$ directly scale on $h$ and so non-dimensional
channel length $L^*=L_p/\xi_0$ is increased. Then, assuming $\xi_0
\propto h$, for $h=1.7$~mm, $h=1.2$~mm and $h=0.8$~mm, one respectively
gets $L^*=98$, $L^*=140$ and $L^*=210$, {\em i.e.}, very extended cells.
As the transverse aspect ratio of the cell also increases when $h$
decreases, we may expect the occurence of 2-D effects
\cite{garchi01c,burmuk01}. In practice, such effects may occur only
below $h_{\rm c}=1.1$~mm and they are negligeable (a slight curvature of
the wave front is observed for $h=0.8$~mm) for the patterns presented in
this paper. Note however, that the smaller $h$ is, the more the system
is sensitive to fluid evaporation.

We have illustrated in the previous section that the dynamics depends on
the distance to $k_{\rm c}$, and we may suspect how variation of $k_{\rm
c}$ as small as one unit of $2\pi/L_{\rm p}$ \cite{tucbar90note} may
modify the dynamics, while the pattern wavenumber has to remains fixed
to satisfy the periodic boundary conditions. For the smallest heights
the uncontrolled variation of $h$ would probably be responsible of a
continuous drift of the Eckhaus stability limit which would thus be
periodically crossed by the system, this process inducing series of
successive wavenumber transitions.

\subsection{Modulations of large period at small $\epsilon$}
\label{sec:MAW_P=L}

\begin{figure}
\begin{center}
\includegraphics[height=6cm]{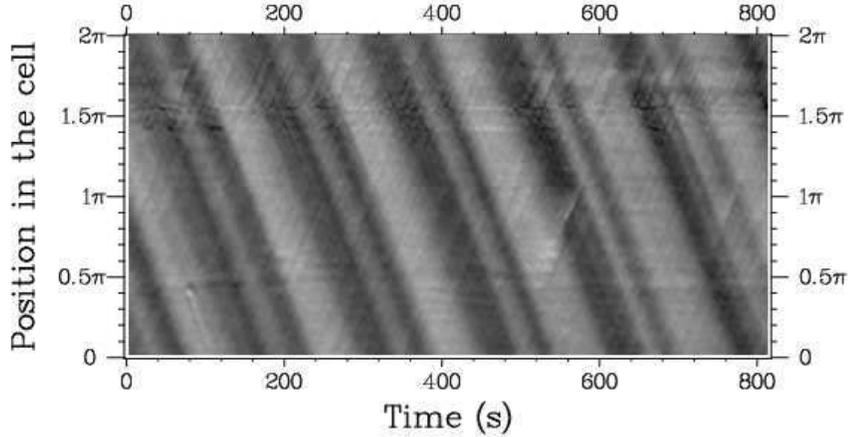}
\end{center}

\caption{Strongly modulated wave at $h=0.80$~mm and $\epsilon=0.25$.
Spatio-temporal diagram of the local wavenumber. This nonlinearly
saturated modulation of a mean wavenumber $k=101(2\pi/L_{\rm p})$
traveling-wave is due to the occurence of the Eckhaus instability. The
instability limit is crossed by varying $k-k_{\rm c}$ at constant
$\epsilon$: a slow evaporation make $k_{\rm c}$ vary while $k$ is kept
constant due to the cell periodicity.}

\label{fig:MAW0.8_im}
\end{figure}

\begin{figure}
\begin{center}
\includegraphics[height=7cm]{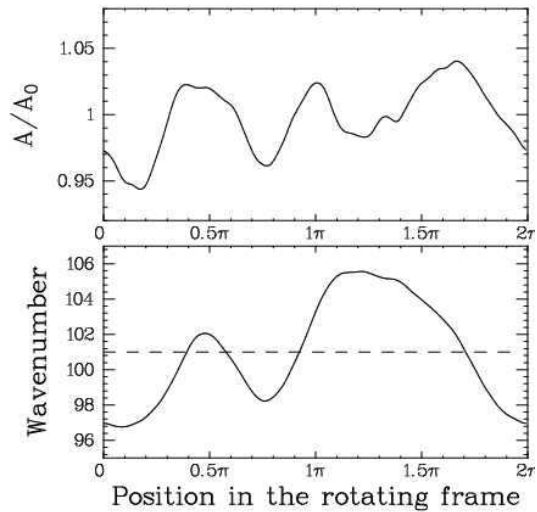}
\end{center}

\caption{Amplitude (top) and wavenumber (bottom) profiles of the
modulated wave presented in Fig.~\ref{fig:MAW0.8_im}. The profiles are
averaged along time in the rotating frame. The amplitude of the
modulation is typically $\pm 5\%$ of the mean value $A_0$ of the
amplitude and of the mean value $k=101(2\pi/L_{\rm p})$ of the
wavenumber (dashed line). }

\label{fig:MAW0.8_prof}
\end{figure}

The following data have been recorded during a long experiment at
constant $\Delta T=4.80$~K and letting $h$ naturally decrease at a rate
of $0.021$~mm/day. The data presented in Fig.~\ref{fig:MAW0.8_im} are
recorded at $h=0.80$~mm, {\em i.e.}, shortly before crossing the
frontier between hydrothermal waves and basic flow (see
Fig.~\ref{StabDiag}). In this region, the slope of the frontier is large
and $\epsilon$ is not precisely known but can we can estimated it to be
roughly $0.25$ ($\Delta T_{\rm c} \simeq 3.9$~K). This experiment
presents thus the classical Eckhaus transition, {\em i.e.}, the crossing
of the Eckhaus boundary at small and decreasing $\epsilon$.
Fig.~\ref{fig:MAW0.8_im} shows the spatio-temporal diagram of the
local-wavenumber for a strongly modulated pattern. The carrier wave has
101 wavelengths in the cell while the wavenumber of the modulation
$K_{\rm M}=2\pi/L_{\rm p}$ is low or its period $P=L_{\rm p}$ is large.
This pattern is saturated by non-linearities, because all quantities
are constant along time. As far as the amplitude of the modulation is
large on all fields, we tried to determine the group velocity and some
CGL coefficients from the frequency-wavenumber and amplitude-wavenumber
relation extracted from local values over the whole data set
\cite{crowil89,burcha99}. We thus get:

\begin{eqnarray*}
&&k_{\rm c}                         = 1.28   \, {\rm mm}^{-1}
                                    = 102.25 \, (2\pi/L_{\rm p})\\
&&\xi_0                             = 3      \, {\rm mm}\\
&&v_G = \partial \omega/ \partial k = 1.61   \, {\rm mm.s}^{-1} 
\end{eqnarray*}

Let us emphasize here that this method is impossible to apply to the
$h=1.7$~mm modulated wave because the modulation amplitude is too tiny.

We also extracted the spatial profiles for the amplitude and
phase-gradients (Fig.~\ref{fig:MAW0.8_prof}). These profiles show the
fine structure of the modulation which is rich in harmonics.

\subsection{Modulations of small period}
\label{sec:tripleMAW}

\begin{figure}
\begin{center} 
\includegraphics[height=5cm]{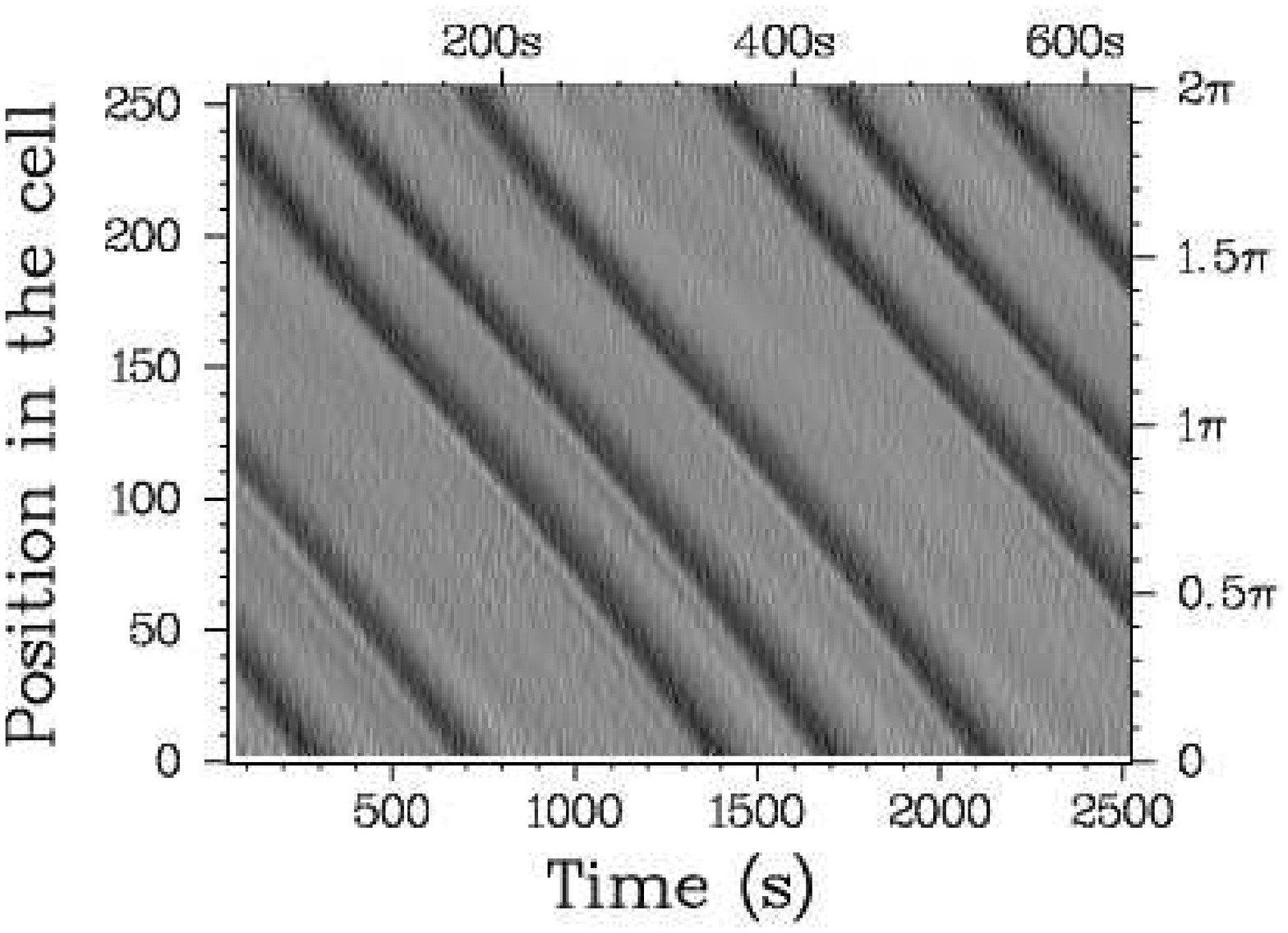}%
\includegraphics[height=5cm]{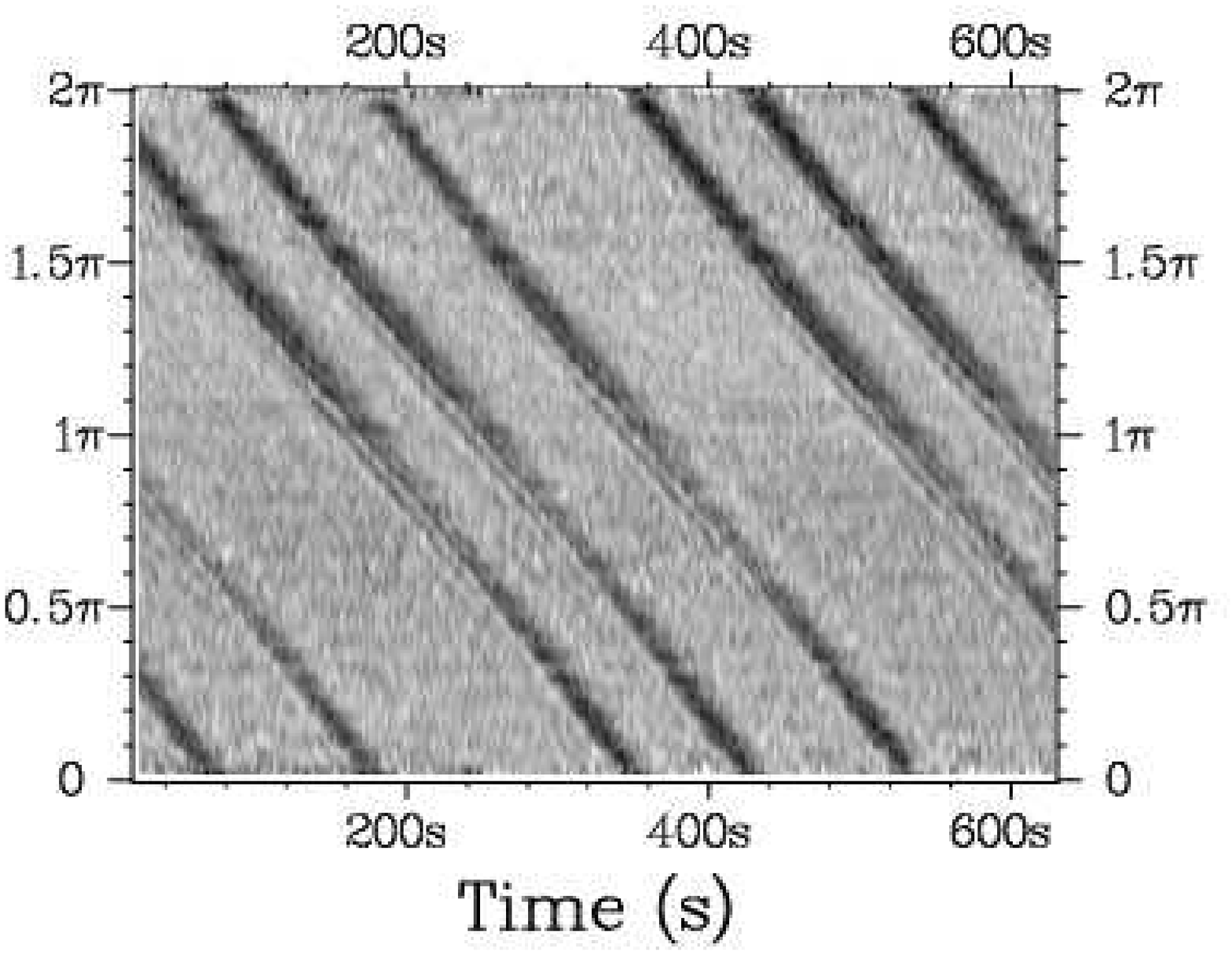} 
\includegraphics[height=4cm]{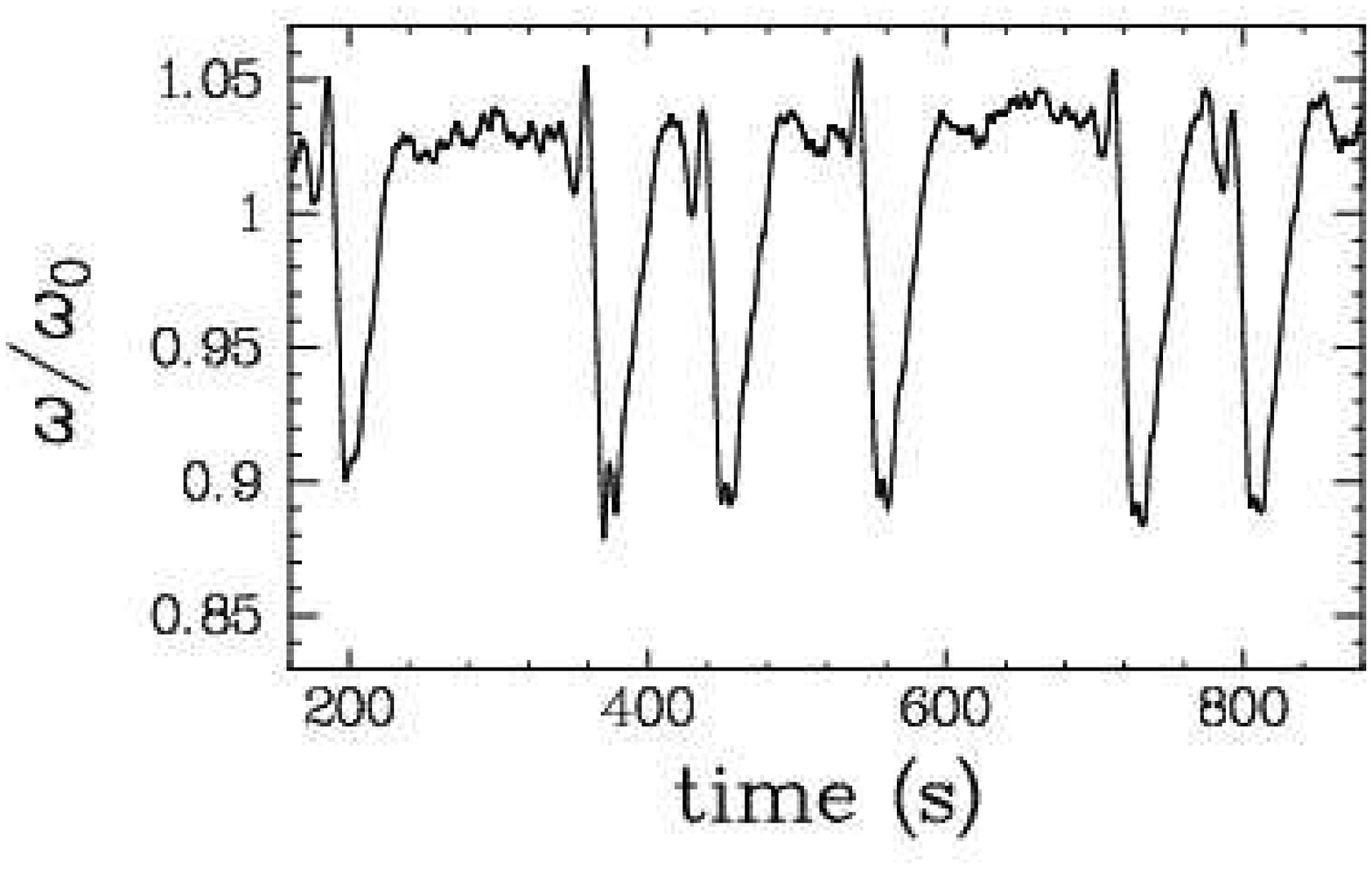}%
\includegraphics[height=4cm]{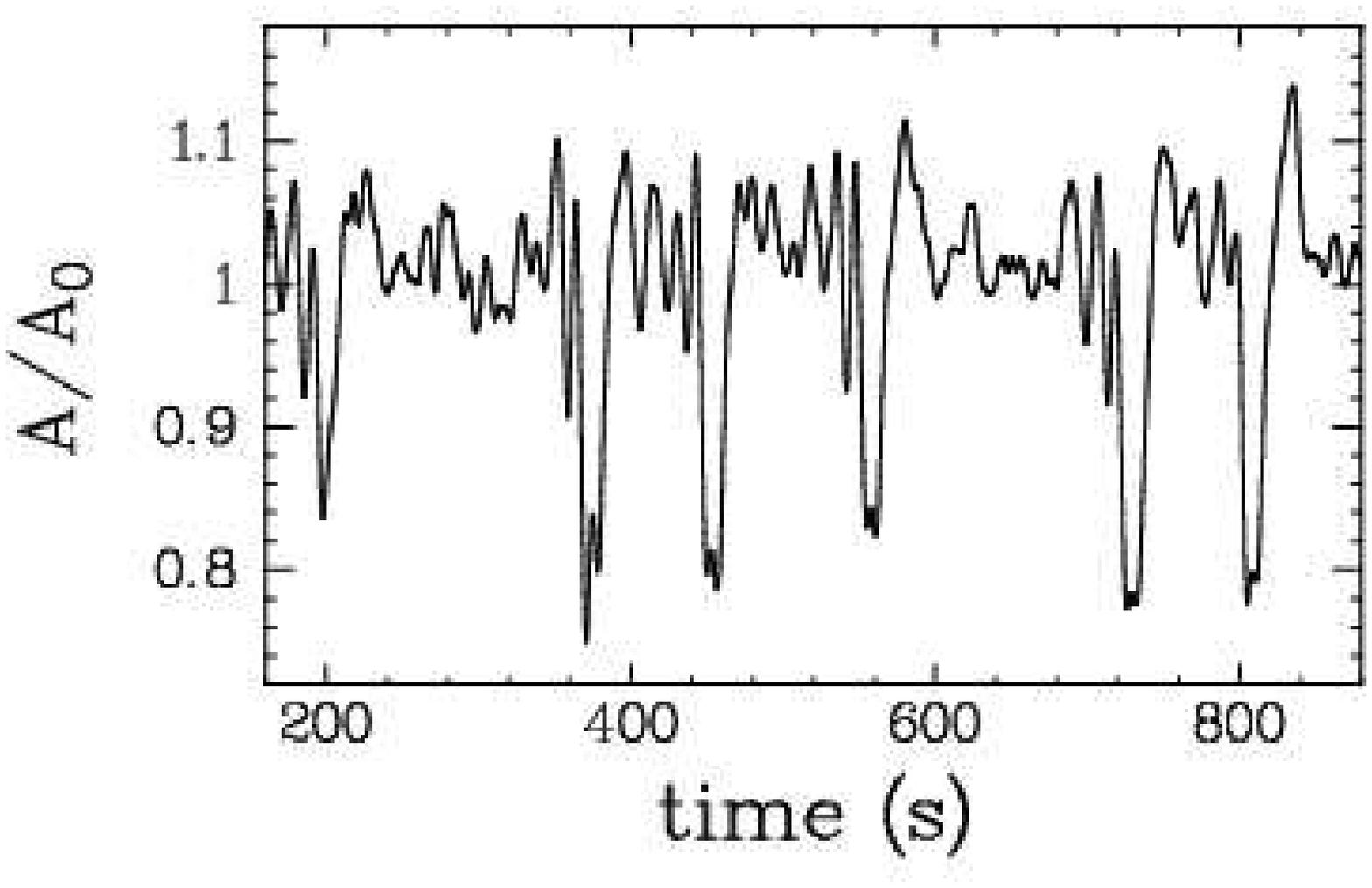} 
\end{center}

\caption{Modulated wave pattern at $h=1.2$~mm and $\Delta T=4.8$~K
($\epsilon \simeq 0.5$). Spatio-temporal diagram and time-serie of the
local frequency are displayed on the left. Spatio-temporal diagram and
time-serie of the local wave-amplitude are displayed on the right. The
maximum local variations are of order of $0.2$ relatively to the mean
values $\omega_0$ and $A_0$. The small oscillations of the modulations
velocity is probably due to a rough adjustment of the channel
horizontality which induces a height modulation.}

\label{fig:tripleMAW_im}
\end{figure}

In another experiment, the fluid height is set to $1.2$~mm and the
temperature difference is increased from the onset value to $\Delta
T=4.8$~K ($\epsilon \simeq 0.5$). Strongly modulated waves are also
observed. Three modulations travel along the cell
(Fig.~\ref{fig:tripleMAW_im}). These modulations have a small spatial
period $P \simeq L/3$ and cannot be described with a simple Fourier
model as Eq.~\ref{eq:defq}. The three modulations look like solitary
waves.

\subsection{Turbulent modulated wave far from threshold}
\label{sec:turbMAW}

\begin{figure}
\begin{center}
\includegraphics[width=14cm]{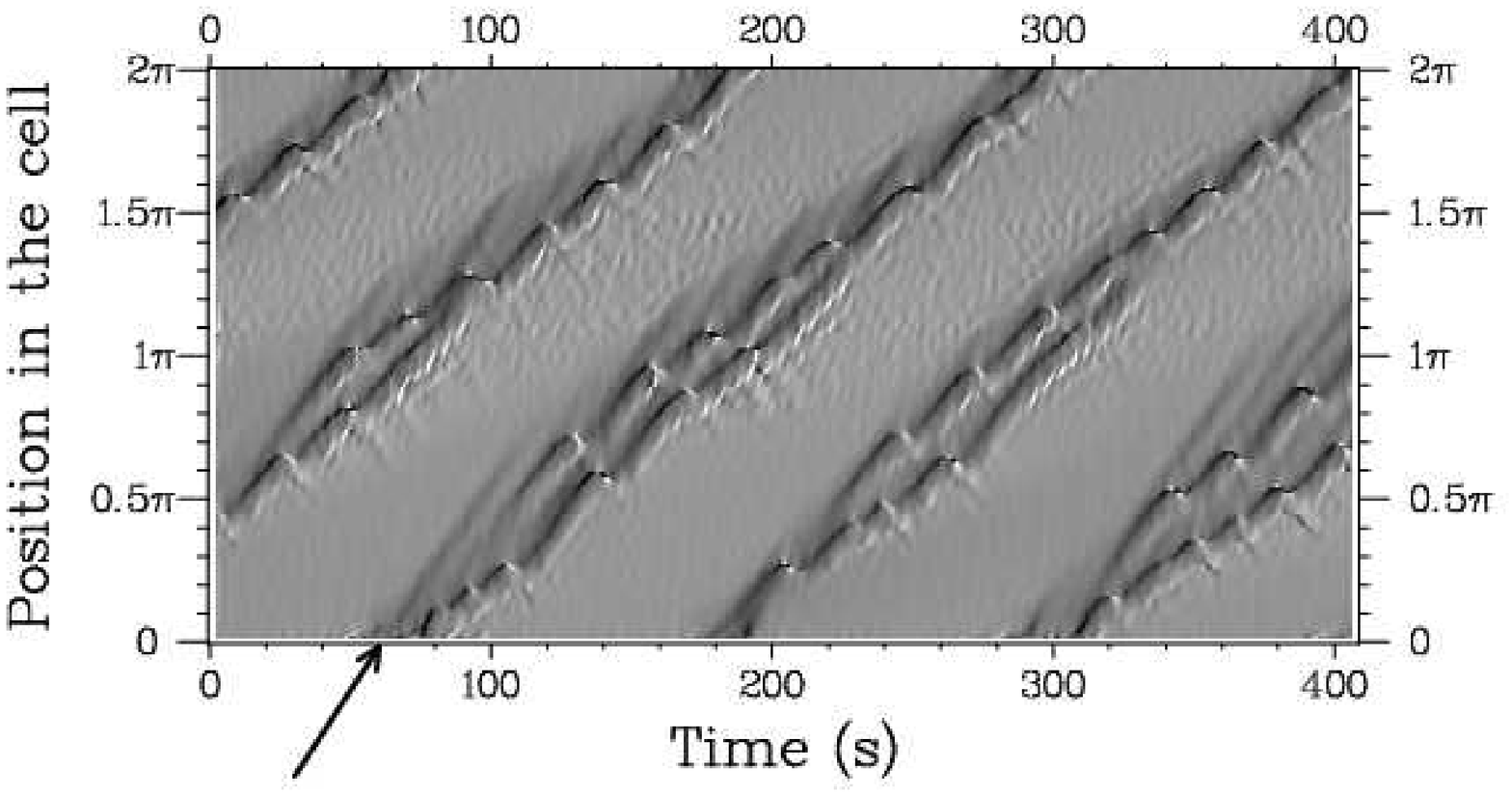} 
\end{center}

\caption{For $h=1.5$~mm and $\Delta T=15.4$K ($\epsilon =4$), two
turbulent modulations are observed to travel in parallel along the
channel. This diagram presents the local wave-frequency. The traveling
modulations produce series of spatio-temporal defects. We may observe
three characteristic velocities: (i) the mean velocity of the global
turbulent pattern, (ii) the velocities of the sharp black or white peaks
corresponding to traveling holes and (iii) the velocity of small smooth
and damped holes or modulations leaving the main pattern, identified by
the arrow, and assumed to represent the effective group velocity. The
mean wavenumber is $k=41(2\pi/L_{\rm p})$ and we note that most holes
are dark, {\em i.e.}, correspond to negative wavenumber peaks. }

\label{fig:turbMAW_im}
\end{figure}

Another type of traveling modulated waves (Fig.~\ref{fig:turbMAW_im})
has been observed for $h=1.5$~mm and $\Delta T=15.4$~K ($\epsilon = 4$).
This pattern of period $P=L/2$ is a turbulent modulation: the wave
amplitude is strongly modulated and often reaches zero in the core of
spatio-temporal defects. These defects occur in the trace of sharp
traveling holes (black or white peaks of frequency on the figure). The
wavenumber is not defined anymore in these spatio-temporal dislocations.
This is an example of defect-mediated turbulence
\cite{cougil89,couelp87} or defect-chaos
\cite{shrpum92,chate94,brutor01a}. One should note that the typical
distance between defects is of order of $L/10$ ($50$~mm) to $L/2$
($250$~mm), whereas the correlation length $\xi=\epsilon^{- 1/2}\xi_0$
is of order of $3$~mm, even smaller than the wavelength $L/41$
($12$~mm).

An interesting observation concerns the local velocity of various
localized patterns of the phase-gradient field: the strongest
modulations, {\em i.e.}, the high peaks of phase-gradients travel more
slowly than smoother phase modulations. Furthermore, all the smooth
modulations are damped and propagate at the same speed, which we
identified as the effective group velocity, represented by an arrow on
Fig.~\ref{fig:turbMAW_im}. This is a general observation for all data in
the annular channel: localized peaks of phase-gradient (close to
space-time dislocations, for example) travel in the laboratory frame at
slower velocity than the group velocity, {\em i.e.}, at a negative
velocity in the comoving frame of the wave pattern, the frame which is
used for theoretical and numerical study of single wave patterns.

\section{Modulated waves in a periodic channel: discussion}
\label{sec:annulus_disc}

\subsection{The context of modulated waves}

We have presented an overview of the known dynamics of modulated waves
in our long annular channel. Modulated waves appear to be everywhere:
near hydrothermal waves onset as well as for high control parameter values,
near the critical wavenumber value as well as far in the side band.

Modulated waves concern mainly single traveling waves, the situation of
most patterns in this paper, but seem to be relevant in competing-wave
patterns as well (Fig.~\ref{fig:turbPS1/2}). Other forms of modulated
waves in bounded boxes are presented in section~\ref{sec:rect_seuil2}
of the companion paper II.

Theoretical study of disordered pattern was mostly initiated on single
1D and 2D Complex Ginzburg-Landau models
\cite{shrpum92,chate94,chaman96}. These first studies where only
concerned by zero mean phase-gradient solutions ($\nu=0$) in periodic
boundary conditions, {\em i.e.}, solutions at $k=k_{\rm c}$. The {\em
mean-phase gradient}, equivalent to our {\em mean wavenumber}, and also
called {\em winding number} is usually defined as:

\begin{equation}
\nu \equiv \frac{1}{2\pi} \int_{0}^{L} \partial_X \Phi dX 
\label{defnu}
\end{equation}

Those works have evidenced the transition from phase- to defect-chaos in
the case of Benjamin-Feir (BF) unstable regimes, and generally ignored
the Eckhaus instability which is unknown at zero mean phase-gradient
unless higher order terms are considered as in section~\ref{sec:HOCGL}.
Our experiment, however, shows supercritical Eckhaus instability regimes
for small mean phase gradients $\nu \simeq 0$. Furthermore, a striking
feature is the overlap of the region of supercritical Eckhaus or
Benjamin-Feir instabilities \cite{janpum92} with the region of phase
chaos at $\nu=0$ \cite{shrpum92}. Both type of solutions correspond to
solutions that may be described by the phase equation ---they may be
called phase-solutions---, and they occur in the same region of the CGL
parameter-space, near the BF border line.

The case of non-zero $\nu$ was pointed out by Montagne {\it et al.}
\cite{monher97} and Torcini {\it et al.} \cite{torfra97} on the basis of
numerical simulations. Recently, this problem has been extensively
studied by numerical analysis based on an equivalent ODE system
\cite{bruzim00}. Both $\nu=0$ and $\nu \ne 0$ cases have been treated by
Brusch and his collaborators in Refs.~\cite{brutor01a} and
\cite{brutor01b} respectively.

Major contributions to the competing ($\lambda > 1$) or cooperating
($\lambda \leq 1$) two-waves problem have been made respectively by van
Hecke {\it et al.} \cite{hecsto99} and Riecke and Kramer
\cite{riekra00}. $\lambda$ is the real part of the coupling coefficient
of CGL Eq.~\ref{eq:cgl}.

Finally, other authors have studied the bounded box case
\cite{tobpro98}, corresponding to the experiments described in II for a
single wave, and the semi-infinite case \cite{coucho99}.

\subsection{Modulated Amplitude Waves --- MAWs}

Modulated wave patterns have been especially recognized by Brusch and
collaborators \cite{bruzim00,brutor01a,brutor01b} who proposed the name
{\em Modulated Amplitude Wave} or MAW, for those solutions of single
CGL equation. All modulated waves described in this section can be
viewed as MAWs. We did not use this vocabulary in the previous sections
in order to avoid possible confusions about both {\em amplitude} and
{\em modulated waves} terms. In modulated {\em amplitude} wave, {\em
amplitude} refers to complex amplitude in CGL, but in experimental data,
the {\em amplitude} refers generally to the modulus of the complex
field. In other word, when discussing above on {\em modulated waves}
(MW), we mean the modulations {\em and} the carrier wave whereas
theoretical MAWs describe the modulation alone, ignoring the carrier
wave. Finally, the {\em group velocity} of MAWs corresponds to the
velocity selection of each particular MAW, while the CGL group velocity
$s$ is eliminated by referential change. However, in the experimental
frame the modulation velocity is, at first order, $s$ (with some HOCGL
corrections), and the specific MAW velocity could be only a small
correction to be extracted from experimental noise. All those reasons
justify the language used in the above sections. Anyway, in this
discussion, observational results and MAWs properties will be directly
compared using MAWs language in order to ease the comparison of
experimental results with theoretical papers. 

MAWs numerical study is based on few parameters: the CGL coefficients
$c_1$ and $c_2$, the mean phase-gradient $\nu$, the spatial period $P$
of the MAW solutions and the size of the box $L$. $P$ and $L$ have been
identically defined above, $\nu$ represents the mean reduced wavenumber
$q=\xi_0(k-k_{\rm c})$ and, as discussed in section~\ref{sec:HOCGL}, our
HOCGL coefficients together with $\epsilon$ play a similar role than CGL
$c_1$ and $c_2$ in parametrizing the problem. Brusch {\it et al.}
generally keep one of the $c_i$ constant (e.g., $c_1=3.5$ in
Ref.~\cite{brutor01b}) and vary the other coefficient to explore the
dynamical regimes and instabilities. This changes the properties of the
system with respect to the Eckhaus/Benjamin-Feir transition and the
$L_1$ and $L_3$ lines. Experimentally, we encountered similar
transitions by simply varying $\epsilon$ which, because of the existence
of HOT, also changes the distance of the system to Eckhaus and BF
transition (section~\ref{sec:model}). On the qualitative point of view,
our exploration of the parameter space is of the same nature.

Let's recall some of the main results of MAWs study. Different types of
MAW solutions have been observed: homoclinic orbits corresponding to
infinite $P$ patterns and heteroclinic orbits corresponding to finite
$P$ patterns. Coherent MAW patterns have been shown to appear through a
forward Hopf bifurcation (HB) and loss their stability through a
saddle-node bifurcation (SN). Therefore, the stable MAW branch is
surrounded by an unstable branch, both connecting at the saddle-node.
Different stable MAW branches may select various group velocity, a
negative velocity branch, a zero velocity branch and a positive velocity
branch through a drift pitchfork bifurcation. Finite size effect have
been shown to be of major importance: transitions may be parametrized by
CGL coefficients as well as by $P$ \cite{brutor01a} and the existence
and the stability of a $P$-period solution strongly depends on the
relative value of $P$ and the box size $L$. Unstable MAWs have been
shown to be the seeds of spatio-temporal defects leading either to
pattern breaks (in the Eckhaus unstable case) or to defect chaos. In
phase chaos regimes, local fluctuations of the phase-gradient or
amplitude can be viewed as single MAWs in interactions.

\subsection{Supercritical Eckhaus modulated wave patterns near $k_{\rm c}$ }
\label{sec:supeck}

It is now well known that supercritical modulations of traveling waves
may result from a supercritical Eckhaus transition
\cite{fauve87,janpum92,mukchi98,zhoouy00,brutor01b}. Since Eckhaus
instability is a low-wavenumber instability, we discussed this behavior
over the amplitude of the first mode $K_{\rm M}=2\pi/L$ \cite{mukchi98}.
However, it is clear (Figs~\ref{fig:mwdamped}, \ref{fig:imsatur} and
\ref{fig:holes}. See also Fig.~3 of Ref.~\cite{mukchi98}) that finite
spatial harmonics contribute to the modulation profile. This shape
(Fig.~\ref{fig:holes_rot}) is qualitatively in very good agreement with
the calculated shape of theoretical MAW solutions. Supercritical Eckhaus
modulations are to be searched among $P=L$ stable MAWs \cite{brutor01b}.

\begin{figure}
\begin{center} 
\includegraphics[width=8cm]{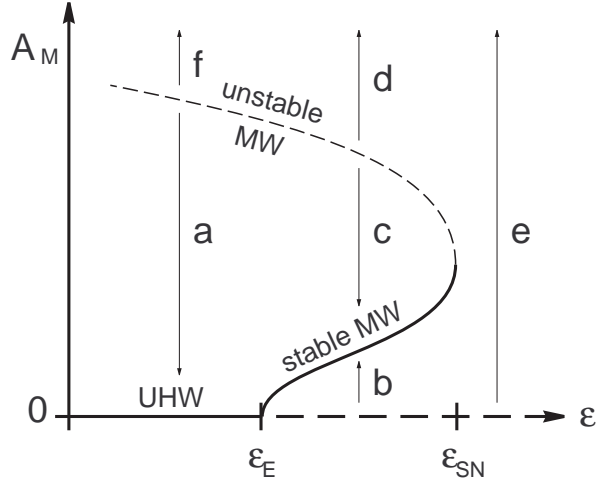} 
\end{center}

\caption{Schematics of the dynamics of experimental modulations. The
control parameter $\epsilon$ is varied in abscissa and a characteristic
instantaneous measure of the modulation amplitude $A_{\rm M}$ (e.g.,
amplitude $|A|$ or phase-gradient variation amplitude) is plotted in
ordinate. Several typical paths (a-f) have been traced. See
section~\ref{sec:supeck} for explanations and discussion. }

\label{fig:supeck}
\end{figure}

Moreover, the temporal dynamic of the experimental modulated wave
solutions is very well illustrated in Fig.~13 of Ref.~\cite{brutor01b}
(on page XXX of the present volume) which presents some spatio-temporal
diagram exhibiting exploding, relaxing and non-linearly saturating MAWs.
Fig.~\ref{fig:supeck} supports an overview of the experimentally
observed modulation regimes inspired by Brusch {\it et al.}'s
presentation. Typical experimental path have been schematized. The main
difference stays in the way a state is prepared: as explained above,
initial modulations are produced by simply changing $\epsilon$
(section~\ref{sec:mwdamped}), so the zero modulation initial condition
is, in practice, unreachable. The dynamics will strongly depend on an
initial hardly controlled condition. Path (a) is the most common
experiment below $\epsilon_{\rm E}$ (Fig.~\ref{fig:mwdamped}), whereas
path (f) illustrates the fragility of Eckhaus stable waves
(section~\ref{sec:mwstable}) and has been the source of strong
questioning about the experiment reproducibility at its beginning! Path
(c) is the usual way for stable modulated patterns to appear because of
the initial amplitude. So path (b) as never been observed and would
probably require a very slow increase of the control parameter to be
produced experimentally. Path (d) is what happens when the control
parameter is increased too much from an initial stable modulated
pattern. Finally, path (e) describes the destabilization of the pattern
(Fig.~\ref{fig:holes}) above $\epsilon_{\rm SN}$. Brusch {\it et al.}
noted that such path, passing close to the saddle-node, seems to
saturate as a stable MAW for a long time before leaving the SN region.
This behavior is clearly observed in the experiment. For example, the
stable modulation in Fig.~\ref{fig:imsatur} is obtained after a long
decrease of an initial modulation as on path (c). Then, at the end of
the acquisition, $\epsilon$ is increased a little. We then observe a
quasi-stable MAW on the first part of Fig.~\ref{fig:holes}. This MAW has
a constant amplitude for its first spatial Fourier mode, but the maximal
value of the phase-gradient slowly increases with time up to $t=870$~s:
the modulation gets sharper and sharper. For $t>870$~s, all Fourier
modes, even the fundamental, start exponentially growing with a short
characteristic time up to the birth of the first spatio-temporal defect.

From a Fourier-modes point of view, Eckhaus
instability appears once the highest unstable wavenumber becomes larger
than $2\pi/L$. Suppose this mode is saturated by the
non-linearities. If the control parameter increases, the spatial
harmonics $2n\pi/L$ also becomes unstable. If higher harmonics are less
saturated than the fundamental mode and the lower harmonics are, the
modulated pattern will become unstable above a second critical control
parameter: the stable-modulation domain is bounded from above.
Such basic Fourier mode description may be explored experimentally. One
can clearly see how, far from $\epsilon_{\rm E}$ as on
Fig.~\ref{fig:mwdamped}, the harmonics decay much faster than the
fundamental. The decay rate ratio between fundamental and first harmonic
is 4.1, close to 4, {\em i.e.}, proportional to $K_{\rm M}^2$. Upper harmonics
remains within the noise level. This effect is in agreement with the
classical low-wavenumber limit of the Eckhaus instability
(Eq.~\ref{eq:sigma}). On the other hand, close to $\epsilon_{\rm E}$,
such comparison seems to fails. Further study would be needed to bring
quantitative conclusions. 

The high degree of similarity between experimental and numerical
realizations allows us to conclude that both experimental modulated
waves and stable MAWs are the actual solutions of supercritical Eckhaus
unstable patterns in periodic boundary conditions traveling-wave
systems. As this does not seem to depend on the exact Ginzburg-Landau
model, it is probably a general property of the Eckhaus modulational
instability regardless of the nature of the wave system.

Janiaud {\it et al.} \cite{janpum92} have calculated the region where
Eckhaus instability is supercritical in the $(c_1,c_2)$ plane. This
region appears to border the BF line $1+ c_1 c_2=0$. We did not
calculate this domain for the HOCGL equation but we noticed that the
$\epsilon_{\rm E}(k)$ curve is quite horizontal ({\em i.e.}, independent of
$k$) in the central region of the stability diagram
(Fig.~\ref{fig:diagstab1}). The parallel is easy to make: Eckhaus
instability becomes Benjamin-Feir instability when all wavenumber are
unstable. So we may imagine a CGL system varying coefficients
$c_1(\epsilon)$ and $c_2(\epsilon)$ such that $1+c_1c_2$ decreases from
positive value to zero at a given $\epsilon_{\rm E}$. This will close
the Eckhaus stable domain from the top with an horizontal tangent at
$k_{\rm c}$. This may be the reason for Eckhaus to be supercritical in
this central band of wavenumber. Otherwise, Brusch {\it et al.}
\cite{brutor01b} also observed the stable MAWs to take place in the
central wavenumber band close to $k_{\rm c}$. Phase-chaos patterns,
which are another type of phase solutions, are also localized in this
central band. More generally, we may propose that phase-solutions
develop on a durable way only in domains of the parameter-space that are
close to the Benjamin-Feir instability limit and for $k$ close to
$k_{\rm c}$.

\subsection{Low wavenumber patterns far from $k_{\rm c}$ }

On the other hand, defect-chaos and subcritical Eckhaus instability are
numerically observed for large mean phase-gradients. Both patterns are
characterized by the presence of defects and are to be studied
considering both phase and amplitude variations \cite{sak90}. For lower
wavenumbers in our experiment, the Eckhaus border line becomes strongly
dependent on $k$. The instability is probably subcritical although this
has not been carefully investigated (section~\ref{sec:square}). Strongly
``square''-modulated patterns (Fig.~\ref{fig:square}) and UHW are
observed at the same region in the stability diagram
(Fig.~\ref{fig:diagstab1}) for $k=45(2\pi/L_{\rm p})$ and
$k=46(2\pi/L_{\rm p})$. This may account for subcritical bistability.
The stability of patterns with wavenumbers outside the central band has
also been investigated with the rectangular channel: the modulational
instability has been carefully investigated for wavenumbers above
$k_{\rm c}$ and is believed to be subcritical at convective threshold as
well as at absolute threshold (section~\ref{sec:rect_seuil2} and
Fig.~\ref{fig:ballon:ann&rect} of~II).

\subsection{Defect chaos patterns far from $k_{\rm c}$ }

For the lowest studied values of $k$ ---the biggest mean phase-gradient
$|q|$ or $|\nu|$---, spatio-temporal defect chaos is observed
(section~\ref{sec:chaos}, Figs~\ref{fig:turbholes} and
\ref{fig:turbPS1/4}). Also, defect chaos is numerically observed far
from the BF line \cite{shrpum92,brutor01a}, and interesting regions of
simultaneous MAWs and defects are observed for large $\nu$. Do
Fig.~\ref{fig:turbholes} data correspond to this region? This question
cannot be answered owing to our present knowledge. 

Another possibility is the occurrence of short-wavelength modulational
instability \cite{matvol93,arakra01}, which may perhaps describe better
the pattern: when the intermittency changes a chaotic state into a
single wave pattern, we again notice this pattern to be modulated at
$K_{\rm M}=2\pi/L_{\rm p}$ with a fast decaying modulation. Thus,
metastable UHW at very low $k$ in the intermittent region seems to be
stable with respect to long-wavelength modulational instability, {\em
i.e.}, the classical Eckhaus instability. Chaotic regimes may thus be
due to short-wavelength instability. A simple way to produce such
instability in our HOCGL model Eq.~(\ref{eq:hocgl}) is to add a second
fifth order term: $|A|^2 A_{XX}$ with a small negative coefficient which
will produce a negative diffusion coefficient for any $k$ at high
wave-amplitude.

The above hypothesis are all based on a single equation model, and may
eventually partly describe the first regime ---a single wave with
traveling modulations and holes (Fig.~\ref{fig:turbholes})--- but is
basically insufficient to describe source/sink patterns
(Fig.~\ref{fig:turbPS1/4}, left). Comparable regimes have been observed
numerically by Riecke and Kramer \cite{riekra00} in the case of small
coupling coefficients $\lambda$. Our estimate for $\lambda = 1.36 \pm
0.2$ (section~\ref{sec:coeffs}) is close to the region where chaotic
competition is shown to occur: the ``domain chaos'' region takes place
between the traveling-wave stability region for $\lambda>\lambda_{\rm
r}$ and the standing-wave stability region for $\lambda<1$. For a given
set of CGL coefficients, $\lambda_{\rm r}$ is found to be $1.3$
\cite{riekra00}. The ``domain chaos'' resembles our observations of
non-symmetrical wave-competition. 

On the other hand, the competing patterns with equivalent ratio of
right- and left-traveling waves (Fig.~\ref{fig:turbPS1/2}, right) may be
directly compared to modulated patterns in the bounded cell described in
II: once source and sink select a zero velocity, they play quite the
same role as boundaries in the rectangular channel \cite{hecsto99}, and
the pattern may be viewed as several adjacent bounded patterns.

Finally, the turbulent traveling modulated pattern
(Fig.~\ref{fig:turbMAW_im}) is an example of defect-chaos pattern
involving a single traveling wave at first order (small
counter-propagating wave patches develop around the defect-cores).

\subsection{Modulation velocities}

A question remains open: what does the modulation velocity
represent? In many case through the paper, we supposed the modulation
velocity to be equivalent to the group velocity. This is true for smooth
modulation, e.g., for the lowest wavenumber mode at $K_{\rm
M}=2\pi/L_{\rm p}$. So this proposition is valid for modulation as on
Fig.~\ref{fig:mwdamped}. This leads to a precise determination of CGL
group velocity $s$ (Eq.~\ref{eq:coeff-s}) by extrapolation at
$\epsilon=0$ for $k \simeq k_{\rm c}$. This value agrees with the value
measured in the rectangular cell (Fig.~\ref{fig:rec:velocity} of II).

In the Eckhaus stable band near $k_{\rm c}$, the modulation velocity
varies linearly (Fig.~\ref{fig:vg_annulus}) as may be predicted by HOCGL
equation (Eq.~\ref{eq:hocgl}). In this frame, we recover Brusch {\it et
al.}'s result: for $\nu=0$, the velocity of MAWs is zero below a drift
pitchfork bifurcation occurring above the Hopf bifurcation to stable
MAWs \cite{brutor01a,brutor01b}. The drift-pitchfork bifurcation has
thus probably be crossed above $\epsilon_{\rm E}$ since the last points
in Fig.~\ref{fig:vg_annulus} show a slightly higher velocity. Anyway,
those measurements are noisy since the drift velocity is typically a
tenth of the group velocity.

Low-$k$ UHW or modulated wave patterns far from $k_{\rm c}$ show a lower
velocity with a similar $\epsilon$-dependence. The reduction of the
velocity may be due to the unfolding of the drift-pitchfork bifurcation
for $\nu \ne 0$ \cite{brutor01b}.

Traveling modulations leading to holes and defects
(Figs~\ref{fig:turbholes} and \ref{fig:turbPS1/4}) as well as traveling
turbulent modulations (Fig.~\ref{fig:turbMAW_im}) exhibit velocity
selection close to defects: the selected velocity is smaller ({\em i.e.},
negative in the rotating frame) than the group velocity and eventually
becomes negative (in the laboratory frame). The velocity varies very
fast near the defect-core. Is this huge effect due to the presence of
counter-propagating waves, or is it due to the proximity of a defect
core?

\section*{Conclusion}

Owing to their apparition via a supercritical instability with finite
frequency, finite wavenumber and finite group velocity, hydrothermal
waves were shown to be very well modeled by an amplitude equation of
the complex Ginzburg-Landau type. Our one-dimensional hydrothermal-wave
system can be considered as an experimental realization of a one-dimensional
system of coupled CGL amplitude equations.

Concerning the amplitude equation itself, we obtained experimental
evidence that higher-order terms should be included. Those terms
probably play an important role for higher values of the control
parameter, but can be discarded close to the onset of the instability. 

In periodic boundary conditions, the Eckhaus secondary instability is
supercritical for wavenumbers close to the critical wavenumber $k_{\rm
c}$, whereas it is rather subcritical far from $k_{\rm c}$. This is
confirmed by observations in non-periodical boundary conditions which
spontaneously selects wavenumbers far from $k_{\rm c}$ and where the
modulational instability is subcritical (paper II).

The development of the secondary Eckhaus modulational instability leads
to the creation of various modulated traveling-wave patterns which have
been presented, illustrated and discussed in the framework of modulated
amplitude waves (MAWs), {\em i.e.}, numerical solutions of the
Eckhaus/Benjamin-Feir unstable CGL equation. The periodic boundary
conditions imposed by the annular geometry favor the emergence of stable
phase solutions, {\em i.e.}, traveling modulations whose amplitude is
stabilized by non-linearities. The amplitude of the phase-gradient
modulation has a very large range: from less than a percent to typically
ten percent of the mean phase-gradient of the carrier wave.

Several examples of defect-chaos have been reported. Eckhaus unstable
patterns nucleate such defects once non-linearly saturated
phase-modulated solutions become unstable. This occurs either because
the control parameter is driven outside the stability domain or because
finite amplitude perturbations break the patterns. Phase patterns have
been shown to be very sensitive and fragile. This fact is by itself a
result of our study and a challenge for experimental work. A major
result provided by the study of MAWs dynamics \cite{brutor01b} is the
fundamental explanation of this fragility: owing to the presence of a
saddle-node bifurcation and thus of an unstable branch over each stable
phase branch, finite perturbations may generate growing phase-gradients
leading to amplitude holes and the breaking of fragile modulated-wave
solutions.

Finally, permanent regimes of defect-chaos have been studied. In most
case they involve both competing right- and left-traveling waves and
appear more complex than the extensively-studied defect-chaos domains
observed for a single complex Ginzburg-Landau model equation. Though
we constructed a higher-order CGL model in order to account for the
reinforcement of the ($x \mapsto -x$) reflection symmetry breaking for
high-amplitude single traveling-wave, we emphasized the restoring ---at
a global level--- of the reflection symmetry in chaotic
patterns of competing right- and left-traveling waves.

\section*{Acknowledgment} 

We wish to thank Joceline Lega, Lutz Brusch, Javier Burguete, Olivier
Dauchot, Martin van Hecke, Chaouqi Misbah, Nathalie Mukolobwiez,
Alessandro Torcini and Laurette Tuckerman for helpful discussions.
Special thanks to Vincent Croquette for providing us his powerful
software {\sc Xvin}. Thanks to Alexis Casner, Thierry Etchebarne and
Nicolas Leprovost who contributed to the data acquisition and
processing, and to C\'ecile Gasquet for her efficient and friendly
technical assistance.

\bibliographystyle{unsrt}	
\bibliography{MWbiblio}		

\begin{thebibliography}{10}
\bibitem{qnu}
The mean wavenumber is also called mean phase gradient and named $\nu$ in the
  MAW literature, \cite{monher97,bruzim00,brutor01a,brutor01b}.
\bibitem{tucbar90note}
The mismatch between the size of a finite periodic box and the closest integer
  number of critical wavelengths and its effect on the {E}ckhaus instability
  has been studied by Tuckerman and Barkley \cite{tucbar90} in the case of
  steady patterns. A fraction of wavelength is shown to have a finite effect.
  As far as we know, no corresponding study for oscillatory patterns have been
  ever performed.
\end{thebibliography}

\end{document}